\documentclass[aps,pra,twocolumn,showpacs,amsmath,amsmath,amssymb,superscriptaddress,longbibliography]{revtex4-1}
\usepackage{amssymb}
\usepackage{graphicx}
\usepackage{dcolumn}
\usepackage{bm}
\usepackage{bbold}
\usepackage{float}

\begin{document}
\author{Ryan Jones}
\affiliation{School of Physics and Astronomy, The University of Nottingham, Nottingham, NG7 2RD, UK}
\affiliation{Centre for the Mathematics and Theoretical Physics of Quantum Non-equilibrium Systems, The University of Nottingham, Nottingham, NG7 2RD, UK}

\author{Jemma A. Needham}
\affiliation{School of Physics and Astronomy, The University of Nottingham, Nottingham, NG7 2RD, UK}
\affiliation{Centre for the Mathematics and Theoretical Physics of Quantum Non-equilibrium Systems, The University of Nottingham, Nottingham, NG7 2RD, UK}

\author{Igor Lesanovsky}
\affiliation{School of Physics and Astronomy, The University of Nottingham, Nottingham, NG7 2RD, UK}
\affiliation{Centre for the Mathematics and Theoretical Physics of Quantum Non-equilibrium Systems, The University of Nottingham, Nottingham, NG7 2RD, UK}

\author{Francesco Intravaia}
\affiliation{Humboldt-Universit\"at zu Berlin, Institut f\"ur Physik, AG
			Theoretische Optik \& Photonik, 12489 Berlin, Germany}
\affiliation{Max-Born-Institut, 12489 Berlin, Germany}

\author{Beatriz Olmos}
\affiliation{School of Physics and Astronomy, The University of Nottingham, Nottingham, NG7 2RD, UK}
\affiliation{Centre for the Mathematics and Theoretical Physics of Quantum Non-equilibrium Systems, The University of Nottingham, Nottingham, NG7 2RD, UK}

\title{Modified dipole-dipole interaction and dissipation in an atomic ensemble near surfaces}
\date{\today}

\begin{abstract}
We study how the radiative properties of a dense ensemble of atoms can be modified when they are placed near or between metallic or dielectric surfaces. If the average separation between the atoms is comparable or smaller than the wavelength of the scattered photons, the coupling to the radiation field induces long-range coherent interactions based on the interatomic exchange of virtual photons. Moreover, the incoherent scattering of photons back to the electromagnetic field is known to be a many-body process, characterized by the appearance of superradiant and subradiant emission modes. By changing the radiation field properties, in this case by considering a layered medium where the atoms are near metallic or dielectric surfaces, these scattering properties can be dramatically modified. We perform a detailed study of these effects, with focus on experimentally relevant parameter regimes. We finish with a specific application in the context of quantum information storage, where the presence of a nearby surface is shown to increase the storage time of an atomic excitation that is transported across a one-dimensional chain.
\end{abstract}


\maketitle

\section{Introduction}

The radiative properties of an emitter can be modified by a change in its environment which alters the structure of the local electromagnetic spectrum. Tailoring the environment in order to obtain desirable radiative properties is a target pursued within a number of scientific disciplines, particularly among those where an efficient detection of fluorescence is relevant (e.g. optical devices, biological imaging) \cite{fort2008}. This idea was first explored in the 1940's by Purcell, who showed that spontaneous atomic decay rates could be modified through coupling of the emitter with a resonant cavity \cite{Purcell1946}. Experimental studies of this effect, which date back to the 1970's, examined how the lifetime of fluorescent $\mathrm{Eu}^{3+}$ ions is modified due to the presence of a metal or dielectric surface \cite{chance1975,amos1997,worthing1999}. In the past few decades, a number of classical and quantum models have been developed to describe this phenomena with a large variety of atomic systems (e.g. NV centers or quantum dots) and materials \cite{kuhn1970,drexhage1970,milonni1973,wylie1984,wylie1985,arnoldus1988,Werra16}. It has been recognized that there are two major contributing factors to the Purcell effect \cite{chance1978}: an interference effect that arises due to the reflection of the emitted field by the surface, and the non-radiative energy transfer from the atom to the surface, which dominates at small atom-surface distances.

A particularly interesting near-field energy transfer process is the decay into surface plasmon polaritons (SPPs). SPPs occur at the boundary between a metal and a dielectric: they are polaritonic states resulting from the interaction of light with the collective oscillations of free electrons at the metal surface \cite{maier2007}. Recently, much work has been focused on creating strong atom-surface coupling using SPPs, since their optical properties can be tailored through the use of different materials and system geometries \cite{chang2007,stehle2014,bartolo2016,dzsotjan2011,javadi2018}. There is also interest in exploiting this strong coupling in order to indirectly couple multiple atoms through a SPP-mediated interaction, particularly in the context of quantum information, where this interaction can be used to generate entanglement between distant qubits \cite{martin-cano2011,hou2014}.

In the many-body context it is known that, if the distance between the atoms is comparable to the atomic emission wavelength, the coupling to the radiation field gives rise to coherent interatomic dipole-dipole interactions. Moreover, the dissipation acquires a collective character in this regime, with the photon emission process being much faster (superradiant) or much slower (subradiant) than the single atom spontaneous decay \cite{dicke1954,agarwal1970,lehmberg1970,james1993}. Lately, this mechanism has been at the center of a growing interest, partly fueled by the recent experimental success in achieving parameter regimes (i.e. dense enough packing of the atoms and/or long wavelength transitions) where the predicted effects are of importance \cite{keaveney2012,Rohlsberger2010,jenkins2016,Bienaime12}.

In free space, the strength of the coherent interaction and the collective decay rate are intrinsically connected, i.e. it is not possible to change one without altering the other. By changing the electromagnetic structure of environment, we can partially overcome this limitation and explore regimes where both coherent and incoherent scattering processes occur at significantly different rates compared to free space values. Although there has been some theoretical studies on the mediation of interactions through surface plasmons \cite{barthes2013,pustovit2009,zhou2011,Palacino17,Sihna18}, several aspects require further investigation. In our paper, we make a detailed analysis of the modified coherent and incoherent interactions due to the presence of metallic and dielectric surfaces. We focus specifically on two situations of experimental relevance: we consider first atoms placed above a single surface \cite{stehle2014,Choquette2010} and then analyze the configuration where they are situated between two identical surfaces \cite{keaveney2012,Rohlsberger2010}. The formalism considered in this paper is, however, powerful enough to account for any configuration involving layered media \cite{Intravaia15a}.

The paper is structured as follows: in Section II, we show how to obtain the master equation which describes the time evolution of the atomic system's density matrix. The parameters describing the collective atomic behaviour are related to the form of the system's Green tensor. We show how the Green tensor can be calculated for both situations of interest, starting from a generic layered medium approach. Finally, we detail the material models used in our work, and provide an explanation of how the surface affects the properties of the electromagnetic field. In Section III, we present a detailed analysis of how the coherent and incoherent interactions in the system are modified due to the presence of the surface(s). Through variation of the atomic transition frequency, the atom-surface distance and the dipole orientation, we identify the wide range of regimes that are possible in this system. In Section IV, we apply this approach to the problem of excitation transport along a one-dimensional chain \cite{olmos2013}, and show how the transport is modified. Section V offers a conclusion and possibilities for future study.

\begin{figure}
\includegraphics[width=\columnwidth]{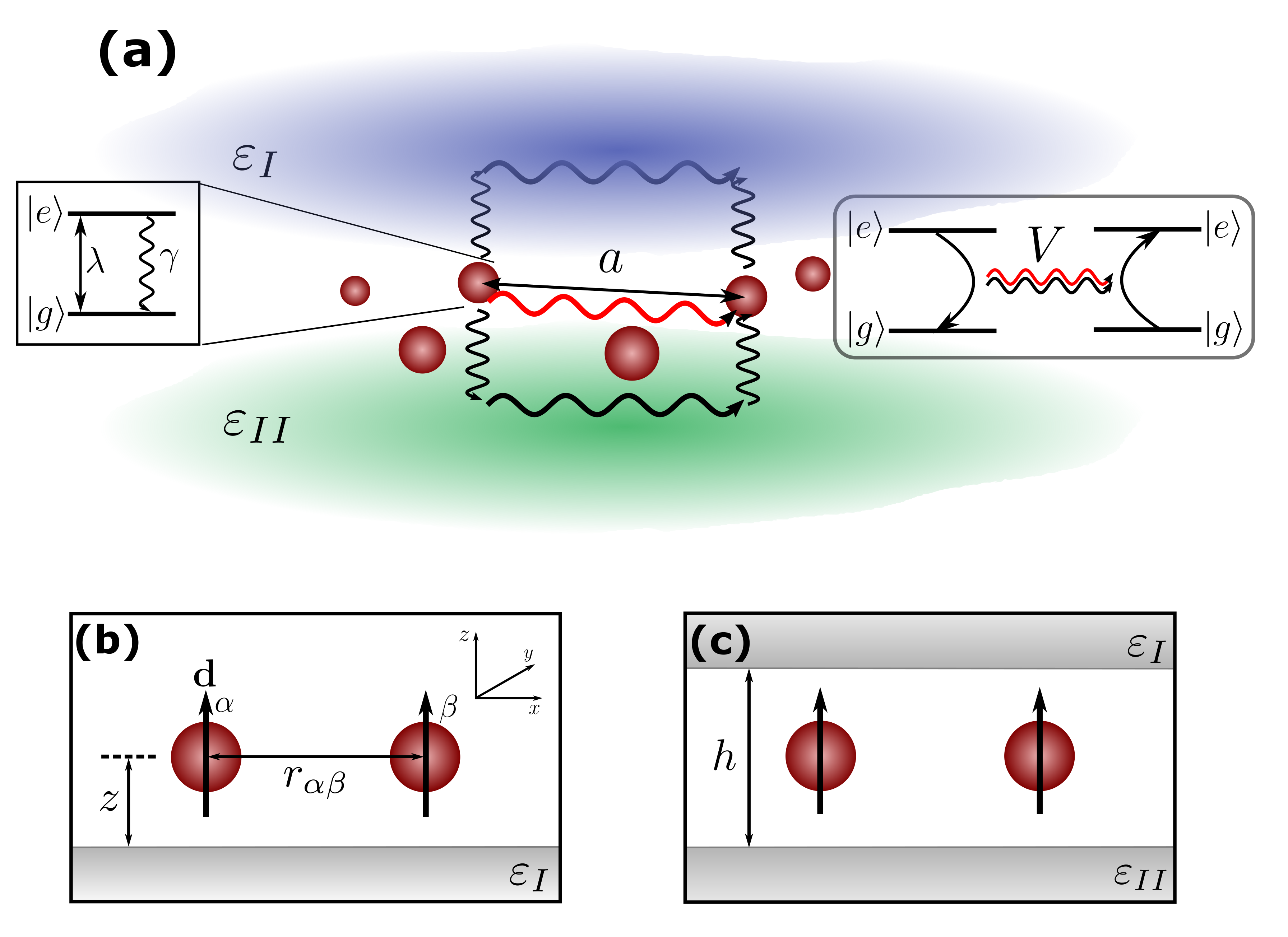}
\caption{\textbf{(a):} A system of $N$ atoms in vacuum is surrounded by one or more layers of metallic or dielectric materials (with dielectric constant $\varepsilon_j$, with $j=I,II,\dots$). Each single atom is described as a two-level system with the excited state $\left|e\right>$ decaying to the ground state $\left|g\right>$ with a free-space rate $\gamma$. As the distance between the atoms $a$ becomes comparable to the wavelength of the $\left|e\right>\to\left|g\right>$ transition $\lambda$, the coherent and incoherent scattering of photons gives rise to long-range exchange interactions between the atoms at a rate $V$. Part of this scattering occurs through coupling to the vacuum radiation field (red) and part through the coupling to the surfaces (black). We study specifically two situations: \textbf{(b):} the atoms at the same distance $z$ from a surface and \textbf{(c):} the atoms placed between two identical surfaces.}\label{fig:system}
\end{figure}

\section{Master equation}

Let us consider $N$ identical two-level atoms placed within a layered medium. We assume that the medium of the layer where the atoms are is the vacuum. The rest of layers are composed by an arbitrary material such as a metal or a dielectric [see Fig. \ref{fig:system}(a)] defined by their (relative) permittivity $\varepsilon_j(\omega)$, where $j=I,II,\dots$ denotes each layer (the layer where the atoms sit has $\varepsilon=1$). The atoms are modeled as two-level systems with ground and excited states labelled $| g \rangle$ and $| e \rangle$ respectively. Each neighbouring pair of atoms is spatially separated by a distance $a$, usually smaller or comparable to the wavelength $\lambda$, characterizing the internal atomic transition energy, i.e. $\hbar \omega_\mathrm{a}=\hbar2\pi c/\lambda$. The atoms have external positions denoted by $\mathbf{r}_\alpha$ ($\alpha = 1,2,...,N$) and the transition dipole moments are denoted by $\mathbf{d}_\alpha$.

Our analysis starts by considering the Hamiltonian that describes a system of $N$ two-level atoms within a \emph{medium-assisted} electromagnetic field at zero temperature. The medium in particular is defined generically for now, in terms of a complex permittivity which can vary spatially. The Hamiltonian reads \cite{perina2001,dung2002,buhmann2004}
\begin{eqnarray*}
{\cal H} &=& \int \mathrm{d^3 \mathbf{r}} \int^{\infty}_0 \!\!\mathrm{d \omega} \hbar \omega \hat{\mathbf{f}}^\dag (\mathbf{r}, \omega) \hat{\mathbf{f}} (\mathbf{r}, \omega) + \sum_{\alpha=1}^N \hbar \omega_\mathrm{a} \hat{\sigma}_{\alpha}^\dag\hat{\sigma}_{\alpha}\\
 &&- \sum_{\alpha=1}^N \int^{\infty}_0 \mathrm{d}\omega [\hat{\mathbf{d}}_\alpha \hat{\mathbf{E}}(\mathbf{r}_\alpha, \omega) + \hat{\mathbf{E}}^\dag(\mathbf{r}_\alpha, \omega) \hat{\mathbf{d}}^\dag_\alpha],
\end{eqnarray*}
where the first two terms represent the matter-electromagnetic field and atoms, respectively, while the last one represents the coupling between the two within the dipole approximation. The quantities $\hat{\mathbf{f}}^{(\dag)}(\mathbf{r}, \omega)$ are polaritonic bosonic operators resulting from Fano-type diagonalisation of the coupled matter-electromagnetic field system and are associated with the annihilation and creation of the corresponding matter-light elementary excitations. Each component $\hat{f}_k(\mathbf{r},\omega)$ of these operators obeys the usual bosonic commutation relations \cite{Scheel08}. Here, the dipole operator reads $\hat{\mathbf{d}}_\alpha = \mathbf{d}_\alpha \hat{\sigma}_\alpha + \mathbf{d}^*_\alpha \hat{\sigma}^\dag_\alpha$ with $\hat{\sigma}_\alpha = |g \rangle _\alpha \langle e|$ being the lowering operator for the $\alpha$-th atom. The electric field operator is defined by
\begin{equation*}
\hat{\mathbf{E}} (\mathbf{r}, \omega) = i \sqrt{\frac{\hbar}{\pi \epsilon_0}} \frac{\omega^2}{c^2}\! \int\! \mathrm{d^3 \mathbf{r}'} \sqrt{\varepsilon_\mathrm{i}  (\mathbf{r}', \omega)} \underline{G}(\mathbf{r},\mathbf{r}',\omega) \hat{\mathbf{f}} (\mathbf{r}', \omega),
\end{equation*}
where $\epsilon_{0}$ is the vacuum (absolute) permittivity. The quantity $\underline{G}(\mathbf{r},\mathbf{r}',\omega)$ represents the electromagnetic Green tensor, which depends on the specific properties of the media under consideration and is detailed in the next section \cite{martin1995}. Finally, $\varepsilon_\mathrm{i}  (\mathbf{r}, \omega)$ is the imaginary part of the permittivity, $\varepsilon(\mathbf{r}, \omega) = \varepsilon_\mathrm{r}(\mathbf{r}, \omega) + i \varepsilon_\mathrm{i}(\mathbf{r}, \omega)$, which satisfies the Kramers-Kronig relations \cite{Jackson75}.

Following standard methods \cite{dung2002,lehmberg1970,james1993}, and under the Born and Markov approximations, one can trace out the electromagnetic environment and obtain a master equation prescribing the dynamics of the internal atomic degrees of freedom through the reduced density matrix $\rho$, which has the form \cite{Breuer02}
\begin{equation} \label{MasterEq}
\dot{\rho} = - \frac{i}{\hbar} [H, \rho] + \mathcal{D}(\rho).
\end{equation}
The first term, determined by the Hamiltonian
\begin{equation} \label{ME-Hamil}
H =-\hbar \sum_{\alpha\neq\beta} V_{\alpha\beta}\sigma^\dag_\alpha \sigma_\beta,
\end{equation}
represents the coherent exchange of excitations among the atoms. Here, $V_{\alpha\beta}$ describes the strength of this exchange interaction and is given by
\begin{equation} \label{eq:VabDef}
V_{\alpha\beta} = \frac{3\gamma\lambda}{2} \hat{d}^* \mathrm{Re}\left[\underline{G}(\mathbf{r}_\alpha, \mathbf{r}_\beta, \omega_\mathrm{a})\right] \hat{d},
\end{equation}
where $\gamma$ is the free-space single atom decay rate \cite{Scheel99}, i.e.
\begin{equation}
\gamma=\frac{\omega_\mathrm{a}^3|\mathbf{d}|^2}{3\pi c^3\epsilon_0\hbar}.
\end{equation}
For simplicity, we have also assumed the all transition dipole moments are aligned, hence $\mathbf{d}_\alpha = \mathbf{d}_\beta = \mathbf{d}=|\mathbf{d}|\hat{d}$.

The second term, describing the dissipative energy transfer processes occurring in the system and representing the collective emission of photons from the system, has the form
\begin{equation} \label{ME-Diss}
\mathcal{D}(\rho) = \sum_{\alpha,\beta} \Gamma_{\alpha\beta} \left( \sigma_\alpha \rho \sigma^\dag_\beta - \frac{1}{2} \{\sigma^\dag_\alpha \sigma_\beta, \rho \} \right),
\end{equation}
where we have defined
\begin{equation} \label{eq:GabDef}
\Gamma_{\alpha\beta} = 3\gamma\lambda\, \hat{d}^* \mathrm{Im}\left[\underline{G}(\mathbf{r}_\alpha,\mathbf{r}_\beta,\omega_\mathrm{a})\right] \hat{d}~.
\end{equation}
The diagonal elements of this matrix, $\Gamma_{\alpha\alpha}$, are the environment-modified single atom decay rates. While these rates do not vary from the free-space value $\Gamma_{\alpha\alpha}=\gamma$ in vacuum, the presence of a nearby surface notably changes their behavior such that $\Gamma_{\alpha\alpha}\neq\gamma$ \cite{chang2007,stehle2014,Choquette2010,bartolo2016,dzsotjan2011,Werra16}. These quantities set the overall time scale at which the dissipation processes become important.

On the other hand, the off-diagonal coefficients $\Gamma_{\alpha\beta}$ determine the collective character of the emission. This can be most easily understood by diagonalising $\Gamma_{\alpha\beta}=\sum_{m}M_{\alpha m}\gamma_m M^\dag_{m\beta}$ such that Eq. \eqref{ME-Diss} becomes
\begin{equation} \label{ME-DissDiag}
\mathcal{D}(\rho) = \sum^N_{m = 1} \gamma_m \left(J_m \rho J^\dag_m - \frac{1}{2} \{ J^\dag_m J_m, \rho \} \right).
\end{equation}
Emission is then described using the collective jump operators $J_m = \sum_\alpha M_{m\alpha}\sigma_\alpha$ and their corresponding decay rates $\gamma_m$. When the atoms are far apart, all off-diagonal elements $\Gamma_{\alpha\beta}\ll\Gamma_{\alpha\alpha}$ and the photons are emitted from the system with a rate close to the single atom one, $\gamma_m\approx\Gamma_{\alpha\alpha}$ $\forall m$. However, as the atoms get closer together $\Gamma_{\alpha\beta}$ become comparable to $\Gamma_{\alpha\alpha}$ and the system features two kind of emission channels: A few with \textit{superradiant} character decaying with rates $\gamma_m\gg\Gamma_{\alpha\alpha}$ and the rest, with \textit{subradiant} character, whose rates are $\gamma_m\ll\Gamma_{\alpha\alpha}$. With this picture in mind, and solely for the purpose of illustration, we will assume in the following that the magnitude of the off-diagonal elements $\Gamma_{\alpha\beta}$ with $\alpha\neq\beta$ represents the collective character of the photon emission.

Finally, note that in this derivation we have implicitly assumed identical transition frequencies $\omega_\mathrm{a}$ for each atom, which may not be a valid assumption when the atoms rest at different heights from the surface \cite{dung2002}. A calculation of the energy shift due to the nearby surface (see Appendix \ref{app:shift} for details) reassures us that for the parameters used in this work not only is the difference between the shift at two different surfaces negligible, but also the overall level shift induced by the medium on each individual atom can be safely neglected, as it is extremely small compared to the atomic transition frequency.

\subsection{Green Tensor in layered media}

The master equation in Eq. \eqref{MasterEq} can include absorption and dispersion effects that arise due to the presence of surfaces through the use of a suitable \emph{Green tensor}. For a layered inhomogeneous medium like the one we are considering here, and assuming all atoms are in the same layer, it can be decomposed as
\begin{equation}
\underline{G}(\mathbf{r}_\alpha,\mathbf{r}_\beta,\omega) = \underline{G}^0(\mathbf{r}_\alpha,\mathbf{r}_\beta,\omega) + \underline{G}^R(\mathbf{r}_\alpha,\mathbf{r}_\beta,\omega),
\end{equation}
where $\underline{G}^0(\mathbf{r}_\alpha,\mathbf{r}_\beta,\omega)$ is a homogeneous (bulk) Green tensor for the material in which the atoms sit (i.e., the vacuum radiation field). The second term $\underline{G}^R(\mathbf{r}_\alpha,\mathbf{r}_\beta,\omega)$ corrects for the inhomogeneities which arise due to the spatial variation of the permittivity and takes into account the electromagnetic scattering at the interfaces.

In the case of atoms in vacuum, the first term has the form \cite{Jackson75}
\begin{equation} \label{BulkGT}
\underline{G}^0(\mathbf{r}_\alpha,\mathbf{r}_\beta,\omega) = \left( \nabla \nabla + k^2 \mathbb{1} \right) \frac{\mathrm{e}^{i k r_{\alpha\beta}}}{4 \pi k^2 r_{\alpha\beta}},
\end{equation}
where $k = \omega/c$, $\mathbf{r}_{\alpha\beta}\equiv \left(x_{\alpha\beta},y_{\alpha\beta},z_{\alpha\beta}\right)=\mathbf{r}_\alpha - \mathbf{r}_\beta$ ($ r_{\alpha\beta}=|\mathbf{r}_\alpha - \mathbf{r}_\beta|$). From this form of the Green tensor one obtains the standard expressions for $V_{\alpha\beta}$ and $\Gamma_{\alpha\beta}$ in vacuum (which we denote from now on $V^{(0)}_{\alpha\beta}$ and $\Gamma^{(0)}_{\alpha\beta}$) that have been obtained in other derivations of the dipole-dipole interaction \cite{lehmberg1970,james1993}
\begin{eqnarray}
V^{(0)}_{\alpha\beta}&=&\frac{3\gamma}{4}\left(\left[1-(\hat{d}\cdot\hat{r}_{\alpha\beta})^2\right]\frac{\cos{\kappa_{\alpha\beta}}}{\kappa_{\alpha\beta}}\right.\\\nonumber
&&\left.-\left[1-3(\hat{d}\cdot\hat{r}_{\alpha\beta})^2\right]\left[\frac{\sin{\kappa_{\alpha\beta}}}{\kappa_{\alpha\beta}^2}+\frac{\cos{\kappa_{\alpha\beta}}}{\kappa_{\alpha\beta}^3}\right]\right)
\end{eqnarray}
and
\begin{eqnarray}
\Gamma^{(0)}_{\alpha\beta}&=&\frac{3\gamma}{2}\left(\left[1-(\hat{d}\cdot\hat{r}_{\alpha\beta})^2\right]\frac{\sin{\kappa_{\alpha\beta}}}{\kappa_{\alpha\beta}}\right.\\\nonumber
&&\left.+\left[1-3(\hat{d}\cdot\hat{r}_{\alpha\beta})^2\right]\left[\frac{\cos{\kappa_{\alpha\beta}}}{\kappa_{\alpha\beta}^2}-\frac{\sin{\kappa_{\alpha\beta}}}{\kappa_{\alpha\beta}^3}\right]\right),
\end{eqnarray}
with the reduced distance $\kappa_{\alpha\beta}=2\pi r_{\alpha\beta}/\lambda=r_{\alpha\beta}\omega_\mathrm{a}/c$.

In order to calculate the reflection term $\underline{G}^R(\mathbf{r}_\alpha,\mathbf{r}_\beta,\omega)$, we use the Green tensor for a generic layered medium, known for example from Refs. \cite{tomas1995, dung2-2002}. It is useful to calculate the tensor for the three-layer case first [Figure \ref{fig:system}(c)], as the tensor for the two-layer case [Figure \ref{fig:system}(b)] follows easily from this result.

We consider the atoms to be in the second layer, which has a thickness $h$. There are two interfaces, with the lower one set to be at $z = 0$ for convenience. For the three-layer geometry the tensor has the form described in Appendix \ref{scatteringtensor}. It is a functional
of $r^q_+$ and $r^q_-$ ($q = s, p$ labels the two polarisations of light), which are the reflection coefficients for the upper and lower interfaces respectively. For local materials these coefficients are given by the Fresnel expressions \cite{Jackson75} and read
\begin{subequations}
\begin{align}
&r^p_+ = \frac{\varepsilon_{I}(\omega) k_{z} -k_{Iz}}{\varepsilon_{I}(\omega) k_{z} +k_{Iz}}, \hspace{0.5cm} r^s_+ = \frac{k_{z} - k_{Iz}}{k_{z} + k_{Iz}},\\
&r^p_- = \frac{\varepsilon_{II}(\omega) k_{z} -k_{IIz}}{\varepsilon_{II}(\omega)k_{z} +k_{IIz}}, \hspace{0.5cm} r^s_- = \frac{k_{z} - k_{IIz}}{k_{z} + k_{IIz}},
\end{align}
\label{eq:FresnelC}
\end{subequations}
where $I$ and $II$ label the upper and lower layer, respectively [see Fig. 1(c)] such that $k_{z}=\sqrt{\omega^{2}/c^{2}-k_{\rho}^2}$ and $k_{iz}=\sqrt{\varepsilon_{i}(\omega)\omega^{2}/c^{2}-k_{\rho}^2}$ ($\mathrm{Im}[k_{iz}]>0$, $\mathrm{Re}[k_{iz}]>0$ and similar for $k_{z}$) for $i=I,II$, and $k_{\rho}=\sqrt{k_{x}^{2}+k_{y}^{2}}$.
To obtain the two-layer case in Figure \ref{fig:system}(b), we simply set $r^q_+ = 0$ in the expression for the Green tensor [see Eq. \eqref{eq:CqDef} in Appendix \ref{scatteringtensor}].

Inserting the full Green tensor into the expressions for $V_{\alpha\beta}$ and $\Gamma_{\alpha\beta}$ reveals that, for a fixed choice of surface material and atomic transition, they depend on three parameters: $(i)$ the separation between the atoms, $(ii)$ their distance from the surface, and $(iii)$ the orientation of their dipoles (relative to both the surface and between each other).

\subsection{Modelling the dielectric constant}

The Fresnel coefficients in Eq. \eqref{eq:FresnelC} depend on the specific form of the dielectric function $\varepsilon(\omega)$ of the different layers. There are several models widely used in the literature to describe different types of materials, some of which are applicable over a wide range of photon energies.

For instance, the Drude model is a simple model used to describe the response of the free electrons in a dissipative metal \cite{maier2007}, such that
\begin{equation}
\varepsilon_D(\omega) = 1 - \frac{\omega^2_p}{\omega^2 + i \omega \gamma_p},
\end{equation}
where $\omega_p$ is the plasma frequency (typically between 5-15 eV) and $\gamma_p$ is a damping frequency (typically between 10 and 100 meV). The model accurately predicts the dielectric constant of metals for photon frequencies that are small compared to the plasma frequency. At higher energies (larger than 1 eV) the contribution to the dielectric constant from bound electrons can no longer be neglected hence the model becomes inaccurate.

The Drude-Lorentz model is an extension of the Drude model which accounts for the effect of bound electrons \cite{powell1970}, such that
\begin{equation}
\varepsilon_{DL}(\omega) = 1 + \sum^b_{j=0} f_j \frac{\omega^2_p}{\omega_j^2 - \omega^2 - i\omega\gamma_j},
\end{equation}
where $\omega_j$ and $\gamma_j$ are the resonance and damping frequencies related to the bound electrons (interband transitions), $b$ is the total number of resonances considered and $f_j$ provides a weighting of the different terms. The term $j = 0$ corresponds to the Drude model, since $\omega_0 = 0$. Throughout this work, we use the Drude-Lorentz parameters from \cite{rakic1998} to model the dielectric constant of silver and other metals.

In order to study the effect of dielectric materials, we choose as an example a glass surface. To describe its dielectric function, we use a modified Lorentz model (see \cite{Djurisic1998} for details). The model uses a dielectric function of the form
\begin{equation}
\varepsilon_L(\omega) = \varepsilon_{\infty} + \sum^b_{j=1} \frac{f_j  \omega^2_j}{\omega^2 - \omega^2_j - i \omega \gamma^{'}_j}.
\end{equation}
where $\varepsilon_\infty$ is a constant representing the value of the dielectric constant as $\omega \rightarrow \infty$, and $\gamma^{'}_j$ is a frequency-dependent damping \cite{Djurisic1998}
\begin{equation}
\gamma^{'}_j = \gamma_j \ \mathrm{exp}\left[-\alpha_{j}\left(\frac{\hbar(\omega-\omega_j)}{\gamma_j}\right)^2\right],
\end{equation}
where $\gamma_j$ describes the damping strength, $\omega_j$ is the oscillator frequency and $\alpha_j$ is an additional (dimensionless) parameter.

\begin{figure}
\includegraphics[width=\columnwidth]{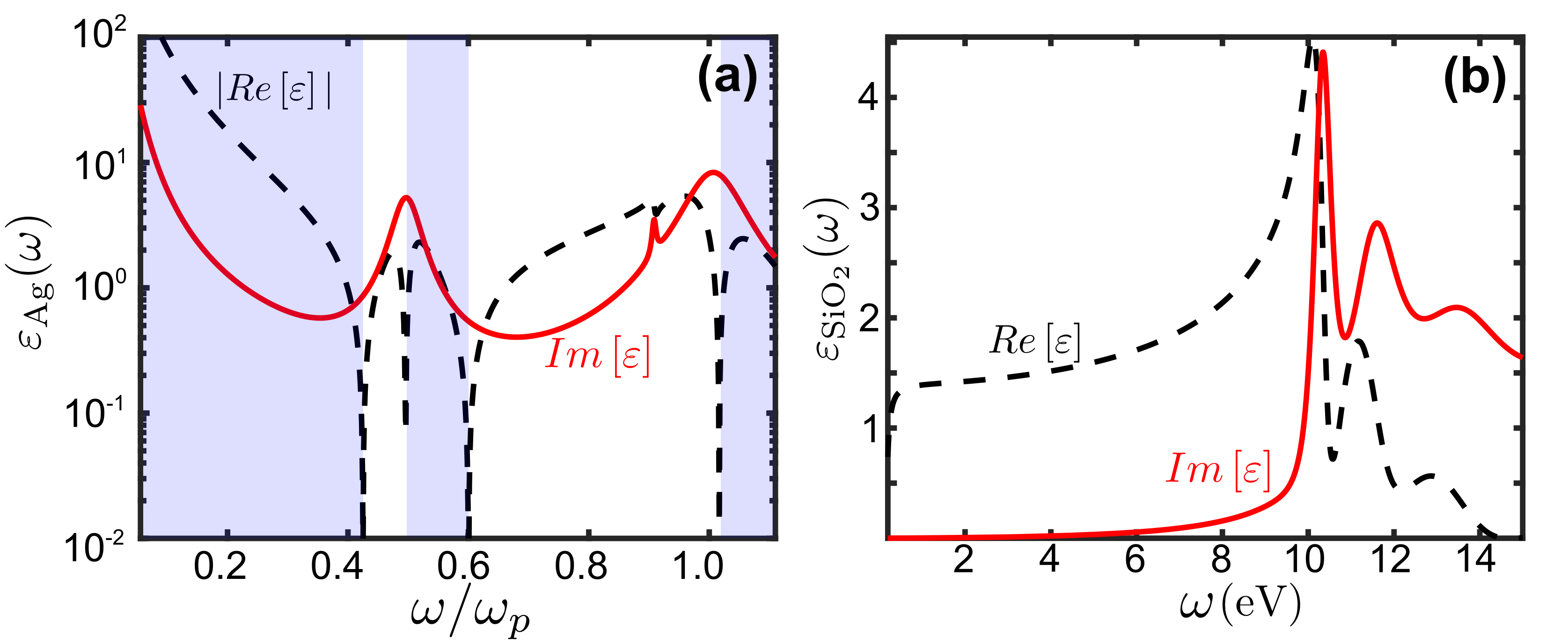}
\caption{Dielectric function for silver \textbf{(a)} and a surface of amorphous SiO${}_2$ (glass) \textbf{(b)}. The shaded areas cover the frequencies where the real part of the dielectric function takes a negative value. \label{fig:eps}}
\end{figure}

In Fig. \ref{fig:eps} the dielectric constants for a metallic surface of silver and a surface of amorphous SiO${}_2$ (glass) are represented as a function of the frequency $\omega$. In the case of silver [Fig. \ref{fig:eps}(a)], the frequency is normalized by the plasma frequency $\omega_p=9.01$ eV, while in the glass [Fig. \ref{fig:eps}(b)] the frequency is simply given in electronvolts. Note that we will be interested in the response of these media at the frequencies corresponding to atomic transitions, which are typically in the range $0.5-5$ eV.

\subsection{Surface interactions}

The presence of the interfaces changes the mode structure of the electromagnetic field, having a number of effects on the atomic interaction and emission.
For atom-surface distances on the order of the transition wavelength and greater, the dominant effect is interference between the emitted and reflected fields. Constructive and destructive interference lead to an increase or decrease in the spontaneous decay rate, respectively.
As the atom-surface distance is reduced, non-radiative processes become more significant. This is due to the increased coupling of evanescent waves from the dipole, whose amplitude decays exponentially with distance from the surface. It is known that this leads to a dramatic increase in the spontaneous decay rate for very small distances \cite{lukosz1977}.

For a metal, these waves can couple with a SPP \cite{Pitarke07}. The existence of such polaritonic surface waves requires a change in sign of the real part of the dielectric function between the two materials. As we are choosing the layer containing the atoms to be vacuum, this condition is generally satisfied for metals at frequencies lower than $\omega_p$ [see Fig. \ref{fig:eps}(a) for silver]. The strength of the coupling to these waves depends on the ratio of the atomic frequency and the plasma frequency of the metal. The maximum coupling is achieved at a resonant frequency which depends on the specific geometry of the system. For a flat metal modeled by the Drude model, and at short distances from the interface (near-field limit) this frequency is around $\omega_p/\sqrt{2}$.
This coupling is interesting as, unlike other near-field effects, which generally only transfer energy to the metal, SPPs can also carry energy between the atoms. For example, an atom can decay and form an SPP which is then reabsorbed by a nearby atom. This therefore provides an alternative mechanism for the dipole-dipole interaction, that is mediated by the surface. This is significant as SPPs can have a long propagation length, so they can be used to couple an ensemble of atoms over longer distances than those achieved using direct radiative coupling alone \cite{pustovit2009}.

In the case of dielectrics, this change of sign of the dielectric constant is typically not present [see, e.g. Fig. \ref{fig:eps}(b) for glass] (when it occurs, one speaks of surface phonon-polariton). Hence, the effect of both the dipole-dipole interactions and the spontaneous decay rate of the atoms near the surface will be expected to be much smaller.

\section{Numerical results}

Here we will investigate numerically how the proximity of surfaces affects the coherent dipole-dipole interactions and the incoherent scattering of photons from a pair of atoms, impacting therefore the collective properties in an ensemble of many atoms. We will evaluate the Green tensor for specific materials (fixing the dielectric function) and vary the remaining parameters: the atomic transition frequency, the distance between the atoms, their distance from the surface, and the relative orientation of the atomic dipole moments with respect to the surface. We analyze the two geometries depicted in Fig. 1(b) and (c), i.e. two atoms with the same dipole orientation, placed at the same distance from a single surface and similarly but with the atoms sitting between two identical surfaces. We focus on silver
and amorphous SiO${}_2$ (glass), although, for completeness, in Appendix \ref{app:materials} we also provide results for other materials.

In all cases, we first evaluate the surface-induced change in the single atom spontaneous decay rate by comparing the diagonal elements $\Gamma_{\alpha\alpha}$ to the free-space value, $\gamma$. We then calculate the surface-induced correction to the dipole-dipole coherent interaction by subtracting the value of the interaction without surface, i.e. $\left(V_{\alpha\beta}-V^{(0)}_{\alpha\beta}\right)/\gamma$. For a better comparison, we also show the values of the interaction without ($V^{(0)}_{\alpha\beta}/\gamma$) and with interfaces ($V_{\alpha\beta}/\Gamma$). Note that in the latter case the rate is scaled by the \emph{modified} single atom decay rate $\Gamma_{\alpha\alpha}=\Gamma$ (identical for both atoms). This is important, as in the presence of a surface it is this quantity and not $\gamma$ that gives a rough idea of how quickly the dissipation kicks in and overcomes any coherent dipole-dipole interactions present in the system. Finally, we will evaluate the value of the off-diagonal terms $\Gamma_{\alpha\beta}$ (normalized again by the modified single atom decay rate $\Gamma$), which represent approximately how collective the process of incoherent emission of photons is.

\subsection{Atoms above a single metal surface}

We calculate here the effect of a metallic silver (Ag) surface on the scattering properties of an ensemble of atoms. As pointed out in the previous section, the main effects of the surface are expected at frequencies smaller than the plasma frequency, here $\omega_p=9.01$ eV, as here the real part of the dielectric constant achieves negative values.

\begin{figure}
\includegraphics[width=\columnwidth]{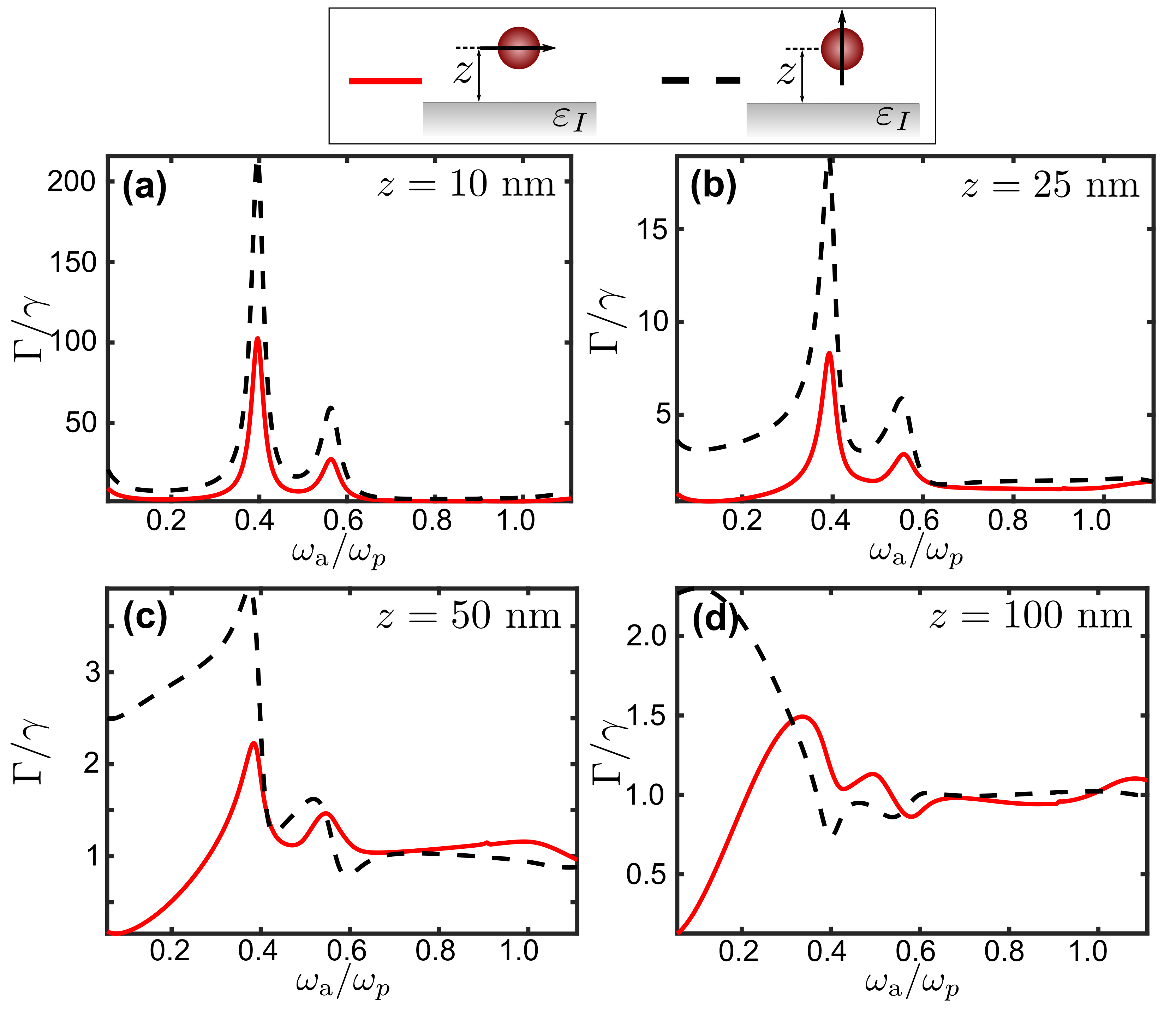}
\caption{Modified single atom decay rate $\Gamma$ as a function of the transition frequency $\omega_\mathrm{a}$. The value in presence of a nearby silver surface, is normalized by the free-space value $\gamma$. The frequency is measured in units of the plasma frequency of the metal ($\omega_p=9.01$ eV here for silver). In each panel the red solid line represents the dipole being parallel to the surface while the black dashed line stands for the dipole being perpendicular to the surface. The atom-surface separations are \textbf{(a)} $z=10$ nm, \textbf{(b)} $z=25$ nm, \textbf{(c)} $z=50$ nm, \textbf{(d)} $z=100$ nm.} \label{fig:RatesAg}
\end{figure}

First, we calculate the effect of the surface in the single atom decay rate of the upper level $\Gamma$ for different values of the transition frequency $\omega_\mathrm{a}$ in units of the surface plasma frequency $\omega_p$. The results are shown in Fig. \ref{fig:RatesAg} for four different values of the distance $z$ of the atom to the surface. One can clearly observe the existence of two main resonances at approximately $\omega_\mathrm{a}/\omega_p=0.4$ and $0.6$ that become broader and less pronounced with increasing distance $z$. These coincide approximately with the frequencies at which the real part of the dielectric constant changes sign. While the latter can be associated with conduction electrons, the former is related to the contribution of bound electrons and interband transitions (more correctly one should also speak in this case of surface phonon-polariton resonances). Due to the stronger interaction, the resonances are more pronounced the smaller the distance to the surface and, for a fixed distance, when the transition dipole moment is oriented perpendicularly to the surface (black dashed lines). Note also that for values of the ratio $\omega_\mathrm{a}/\omega_p\sim 10^{-1}$ the decay rate of an atom with a transition dipole moment oriented parallel to the interface is always reduced with respect to the free-space value $\gamma$, even achieving values very close to zero. This behavior can be understood as resulting form interference effects and compensation between the two polarizations of light. Indeed it does not occur when the dipole is orthogonal to the surface, where only the $p$-polarization is relevant.

\begin{figure}
\includegraphics[width=\columnwidth]{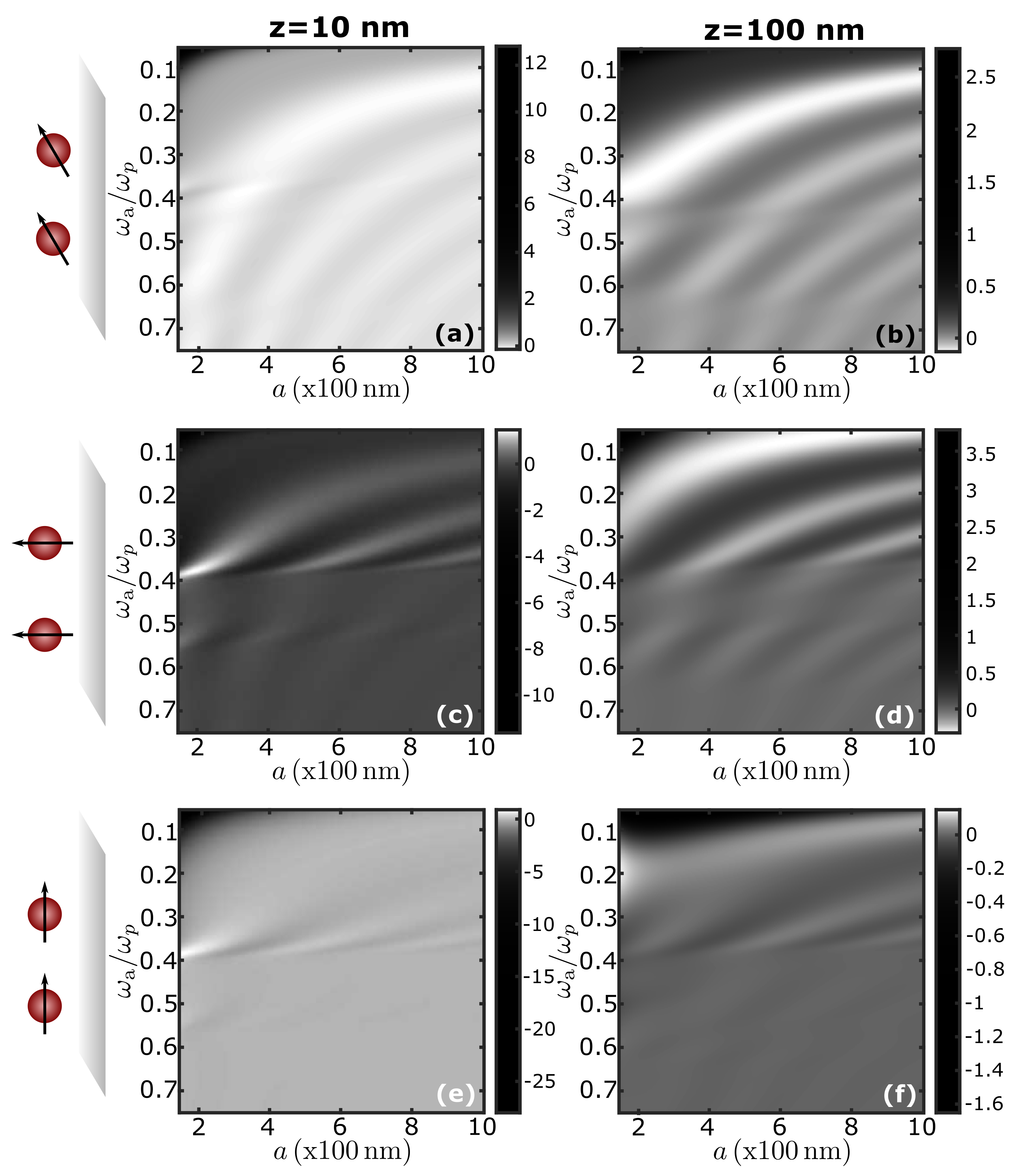}
\caption{Correction to the dipole-dipole coherent interaction due to the silver surface, $(V_{\alpha\beta}-V_{\alpha\beta}^{(0)})/\gamma$ as a function of the distance between the two atoms $a$ and the frequency $\omega_\mathrm{a}$ of the transition $\left|e\right>\to\left|g\right>$. We use the same parameters of Fig. \ref{fig:RatesAg}. The panels \textbf{(a)} and \textbf{(b)} describe the situation where the dipoles are oriented parallel to each other and to the surface. In the panels \textbf{(c)} and \textbf{(d)} the dipole vectors are orthogonal to the surface. In \textbf{(e)} and \textbf{(f)} the dipoles are instead aligned and parallel to the surface. Both atoms are at the same distance from the surface:  $z=10$ nm for \textbf{(a)}, \textbf{(c)} and \textbf{(e)}; $z=100$ nm for \textbf{(b)}, \textbf{(d)} and \textbf{(f)}.}\label{fig:VAg2D}
\end{figure}

Let us consider the contribution to the dipole-dipole interaction solely due to the interaction with the surface, $(V_{\alpha\beta}-V_{\alpha\beta}^{(0)})/\gamma$. We show the results in Fig. \ref{fig:VAg2D} for a fixed value of the distance of the dipoles to the surface ($z=10$ and $100$ nm) and for three orientations of the dipole moments with respect to each other and the interface. Here, clear changes can be observed at the same values of the frequencies where the single atom decay rates $\Gamma$ were enhanced and, correspondingly, the real part of the dielectric constant changes sign. We also observe that the effect of the surface on the interaction is typically larger for small values of the atomic frequencies, relative to the plasma frequency $\omega_p$, and also generally larger for small values of the distance $z$ of the atoms, which seems intuitive as for smaller distances one would expect a larger coupling of both systems.

\begin{figure}
\includegraphics[width=\columnwidth]{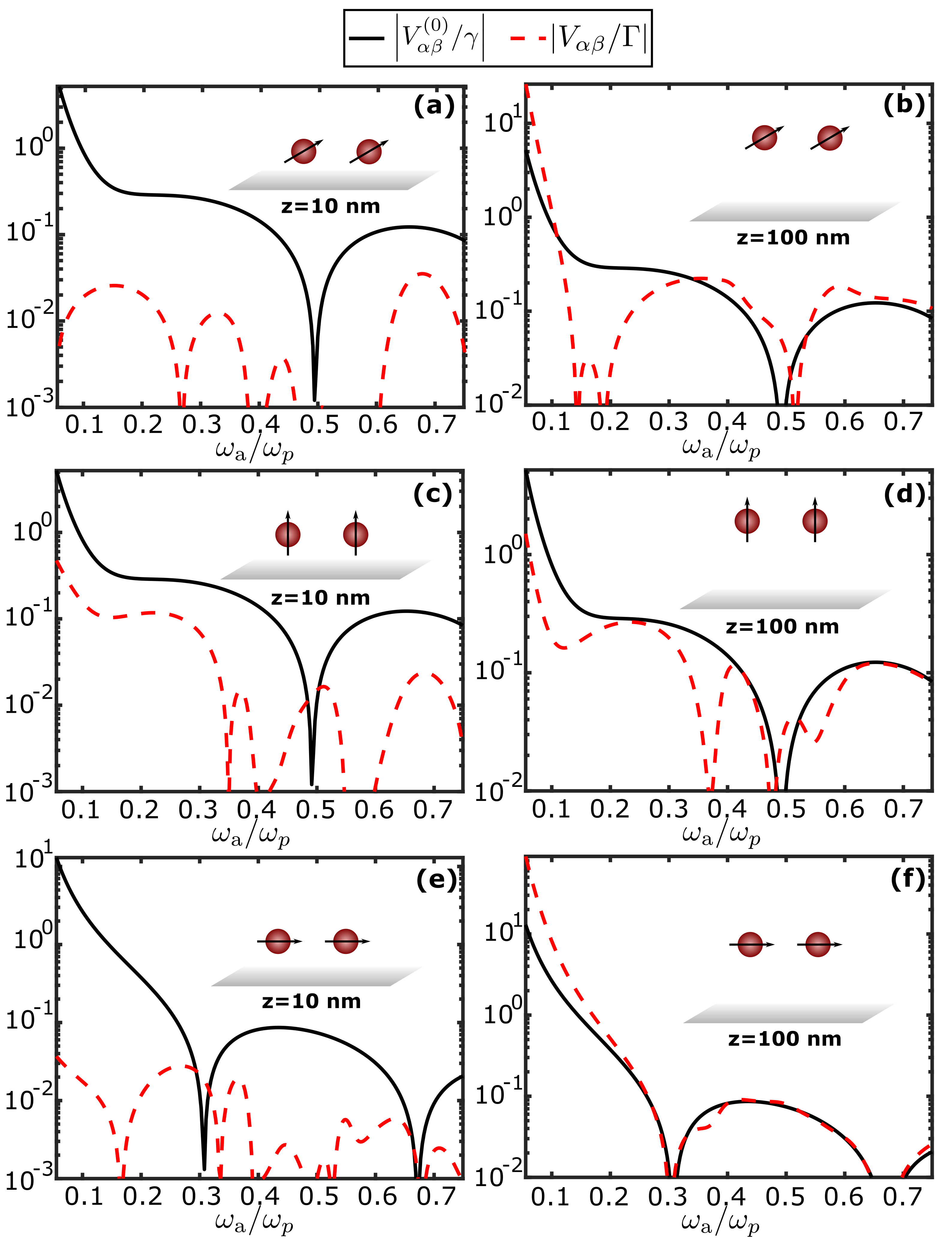}
\caption{Coherent dipole-dipole interaction in free-space, $V_{\alpha\beta}^{(0)}/\gamma$ (black solid line), and in the proximity of a surface, $V_{\alpha\beta}/\Gamma$ (red dashed line), as a function of the frequency $\omega_\mathrm{a}$. The two dipoles are at a fixed distance $a=200$ nm. The surface is silver. In the panels \textbf{(a)} and \textbf{(b)} the dipoles are parallel to each other and to the surface. In \textbf{(c):} and \textbf{(d):} they are oriented orthogonal to the surface. In \textbf{(e):} and \textbf{(f):} the dipole vectors are aligned and parallel to the surface. The distance from the surface for both atoms are: $z=10$ nm in \textbf{(a)}, \textbf{(c)} and $z=100$ nm in \textbf{(e)}, and \textbf{(b)}, \textbf{(d)} and \textbf{(f)}.} \label{fig:VAg1D}
\end{figure}

To get a better understanding of how the interactions are affected by the presence of the surface, we calculate in Fig. \ref{fig:VAg1D} the value of the total interaction in the absence (black solid lines) and in the presence (red dashed lines) of a nearby surface at a fixed distance between the atoms (here $a=200$ nm). As explained above, in the latter we divide the value of $V_{\alpha\beta}$ by the modified single atom decay rate $\Gamma$, as this will better indicate if coherent exchange can indeed be observed in such a system before the dissipation (decay) kicks in. Interestingly, we observe that for all dipole orientations considered the effect of the surface at very short distances (showing $z=10$ nm) is to bring the ratio between the coherent interactions and the dissipation very close to zero ($V_{\alpha\beta}/\Gamma\to 0$ for all values of the frequency $\omega_\mathrm{a}$). For larger distances from the surface (here we show $z=100$ nm) the action of the surface is almost negligible for values of the atomic frequency comparable to $\omega_p$. However, for small values of $\omega_\mathrm{a}/\omega_p$ the coherent interaction $V_{\alpha\beta}/\Gamma$ can achieve much larger values than the free-space ones $V_{\alpha\beta}^{(0)}/\gamma$.

\begin{figure}
\includegraphics[width=\columnwidth]{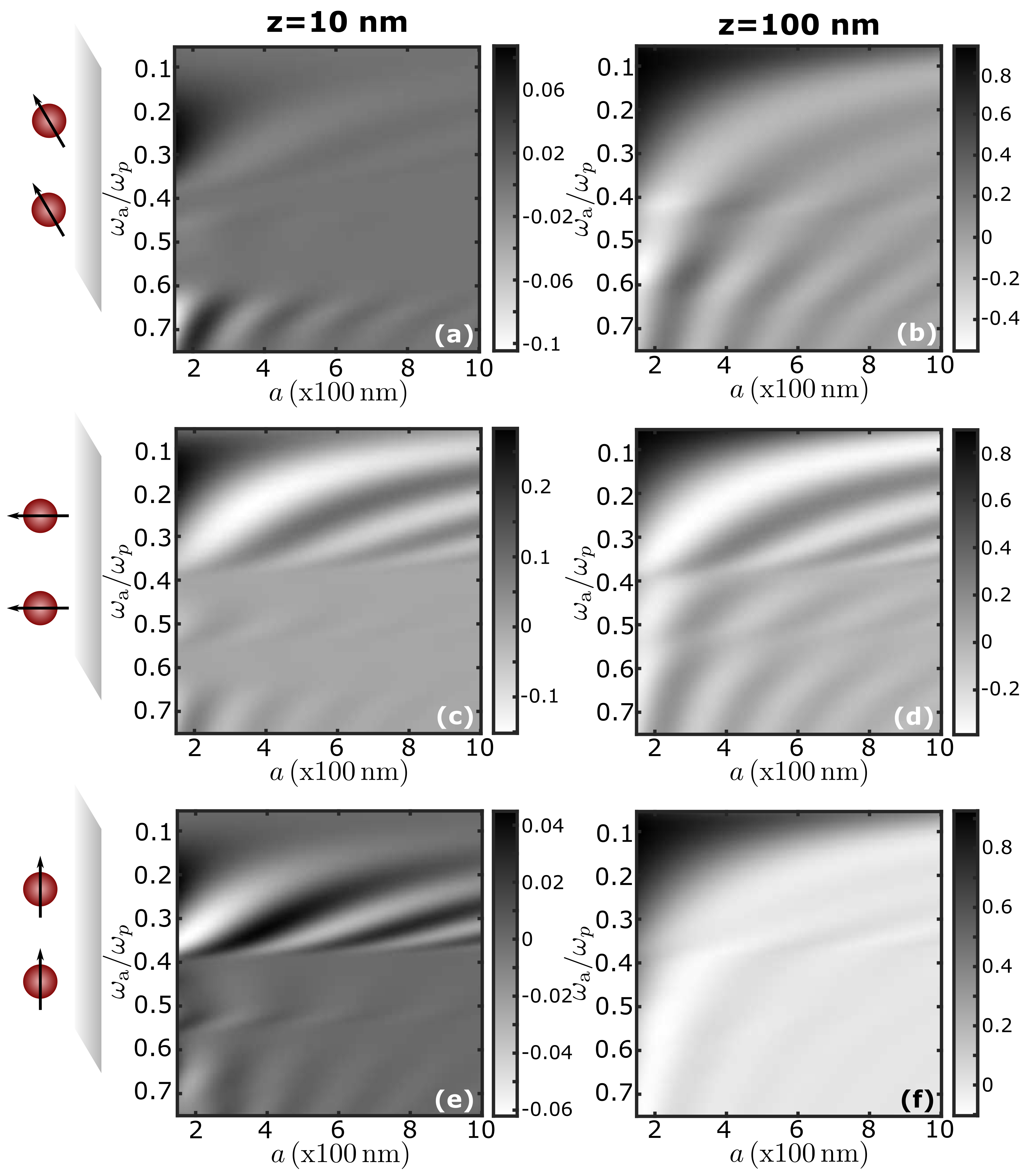}
\caption{Off-diagonal element $\Gamma_{\alpha\beta}$ divided by the modified single atom decay rate $\Gamma$ as a function of the distance between the two atoms $a$ and the frequency $\omega_\mathrm{a}$. In the panels \textbf{(a)} and \textbf{(b)} the dipoles are parallel to each other and to the surface. In \textbf{(c):} and \textbf{(d):} they are oriented orthogonal to the surface. In \textbf{(e):} and \textbf{(f):} the dipole vectors are aligned and parallel to the surface. The atoms are at the same distance from the surface: for \textbf{(a)}, \textbf{(c)} and \textbf{(e)} $z=10$ nm, and for \textbf{(b)}, \textbf{(d)} and \textbf{(f)} $z=100$ nm.} \label{fig:GAg2D}
\end{figure}

\begin{figure}
\includegraphics[width=\columnwidth]{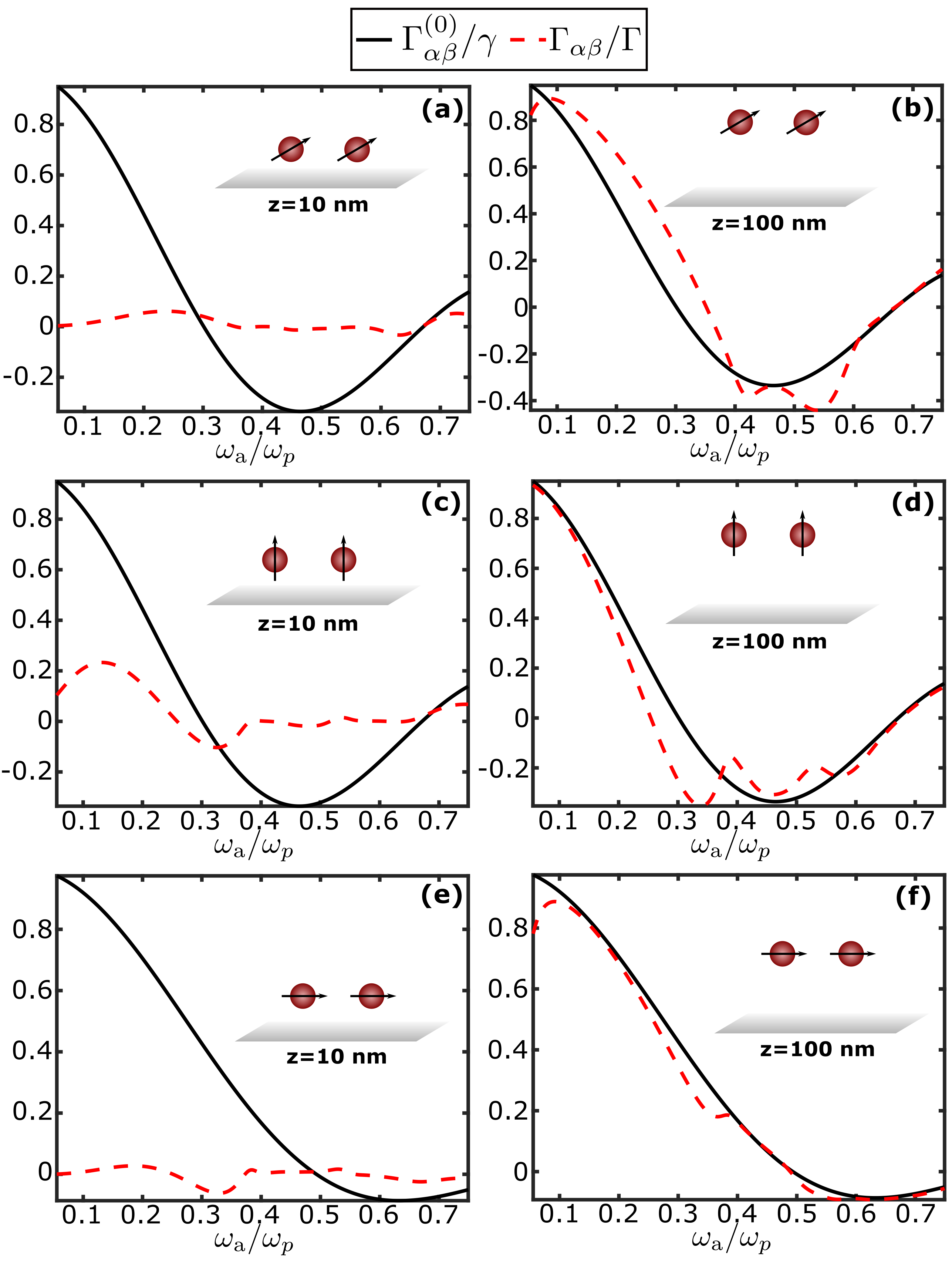}
\caption{Off-diagonal element of the incoherent scattering matrix in free-space $\Gamma_{\alpha\beta}^{(0)}/\gamma$ (black solid line) and in the vicinity of a surface $\Gamma_{\alpha\beta}/\Gamma$ (red dashed line), as a function of the frequency $\omega_\mathrm{a}$. The two dipoles are at a fixed distance $a=200$ nm. In the panels \textbf{(a)} and \textbf{(b)} the dipoles are parallel to each other and to the surface. In \textbf{(c):} and \textbf{(d):} they are oriented orthogonal to the surface. In \textbf{(e):} and \textbf{(f):} the dipole vectors are aligned and parallel to the surface. The atoms are at the same distance from the surface: for \textbf{(a)}, \textbf{(c)} and \textbf{(e)} $z=10$ nm, and for \textbf{(b)}, \textbf{(d)} and \textbf{(f)} $z=100$ nm.} \label{fig:GAg1D}
\end{figure}

To characterize how the collective character of the dissipation is affected by the presence of a nearby surface, in Fig. \ref{fig:GAg2D} we show the value of the off-diagonal element $\Gamma_{\alpha\beta}/\Gamma$ as a function of the distance between the corresponding two atoms $a$ and the frequency of the atomic transition considered. For atoms placed very close to the surface ($z=10$ nm) again the value of $\Gamma_{\alpha\beta}/\Gamma$ is dramatically reduced with respect to its value in vacuum (as can be better observed in Fig. \ref{fig:GAg1D} for a fixed distance between the atoms $a=200$ nm). This in turn means that the decay of the atoms loses its collective character, as the diagonal elements $\Gamma$ are much larger than the off-diagonal ones: super- and sub-radiant decay disappears. On the other hand, at larger distances from the surface ($z=100$ nm) the ratio between off-diagonal and diagonal elements in the dissipation matrix changes only very slightly with respect to the surface-free case.

The previous results show that, when the atoms are very close to a metal surface (here we show the results for $z=10$ nm and silver, but the results are qualitatively similar for a range of distances and different metals, as shown in Appendix \ref{app:materials}) the collective characteristics that arise in dense atomic gases such as dipole-dipole interactions and collective decay of excitations (super- and subradiant states) are suppressed. This is mainly due to the strong coupling to the radiation modes of the surface, which makes the decay rate of each atom individually increase dramatically (see, e.g., Fig. \ref{fig:RatesAg}). A much richer situation arises when the atoms are at intermediate distances from the surface, e.g. $z=100$ nm. Here, depending on the orientation of the dipoles and the frequency of the atomic transition, one can modify the coherent dipole-dipole interaction to make it much stronger or weaker than in the surface-free case. In particular, this occurs for values of the frequency much smaller than the surface plasmon frequency $\omega_p$, which for example is the case for typical atomic frequencies $\omega_\mathrm{a}\approx 0.5-2.5$ eV (in silver $\omega_p=9.01$ eV). Physically, this outcome can be understood as resulting from a weaker single atom dissipation due to interference effects, compensation between the two polarizations, and reduction of the atom-surface coupling strength (proportional to the dissipation in the metal and to the ratio $\omega_\mathrm{a}/\omega_{p}$).

\subsection{Atoms above a dielectric surface}

\begin{figure}
\includegraphics[width=\columnwidth]{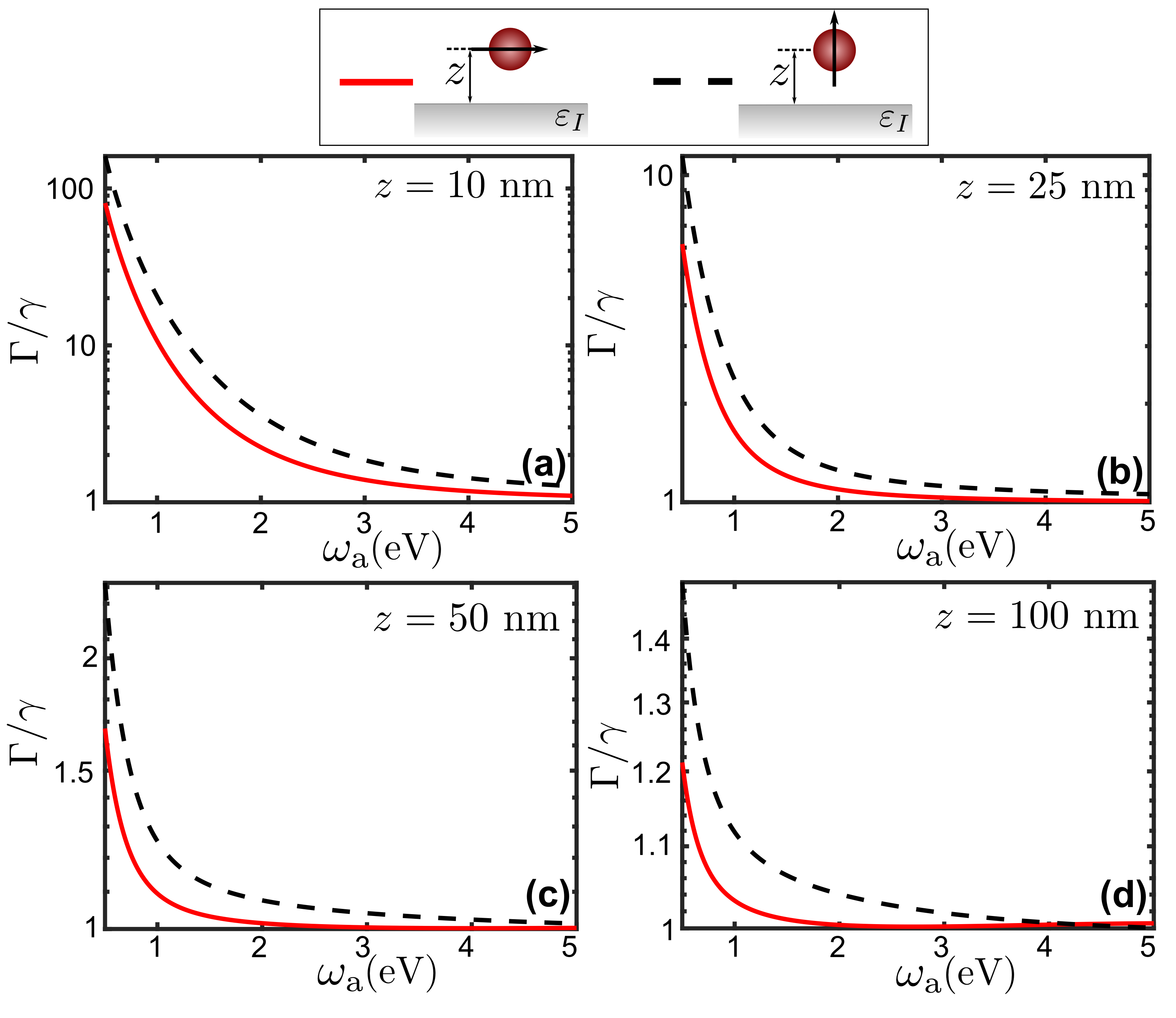}
\caption{Modification in single atom decay rate $\Gamma$ due to the presence of a nearby surface of glass. The value is represented as a function of the frequency $\omega_\mathrm{a}$ and is normalized to the free-space decay rate $\gamma$. In each panel the red solid line represents the situation where the dipole is parallel to the surface while the black dashed line stands for the dipole being perpendicular to the surface. The atom is at a distance \textbf{(a)} $z=10$ nm, \textbf{(b)} $z=25$ nm, \textbf{(c)} $z=50$ nm, \textbf{(d)} $z=100$ nm from the surface.}\label{fig:RatesGlass}
\end{figure}

The interactions with a dielectric surface are expected to have a less prominent effect on the scattering properties of the system, as the dielectric constant of the medium does not change sign at the interface for any value of the atomic frequency $\omega_\mathrm{a}$ [see Fig. \ref{fig:eps}(b)]. Indeed, one can observe this already in Fig. \ref{fig:RatesGlass}, where the modified decay rates $\Gamma$ of a single atom at different distances from a surface of glass are displayed. The modified rate is always larger than the free-space one, $\gamma$, and the changes are only appreciable for very low values of the atomic frequency $\omega_\mathrm{a}$.

\begin{figure}
\includegraphics[width=\columnwidth]{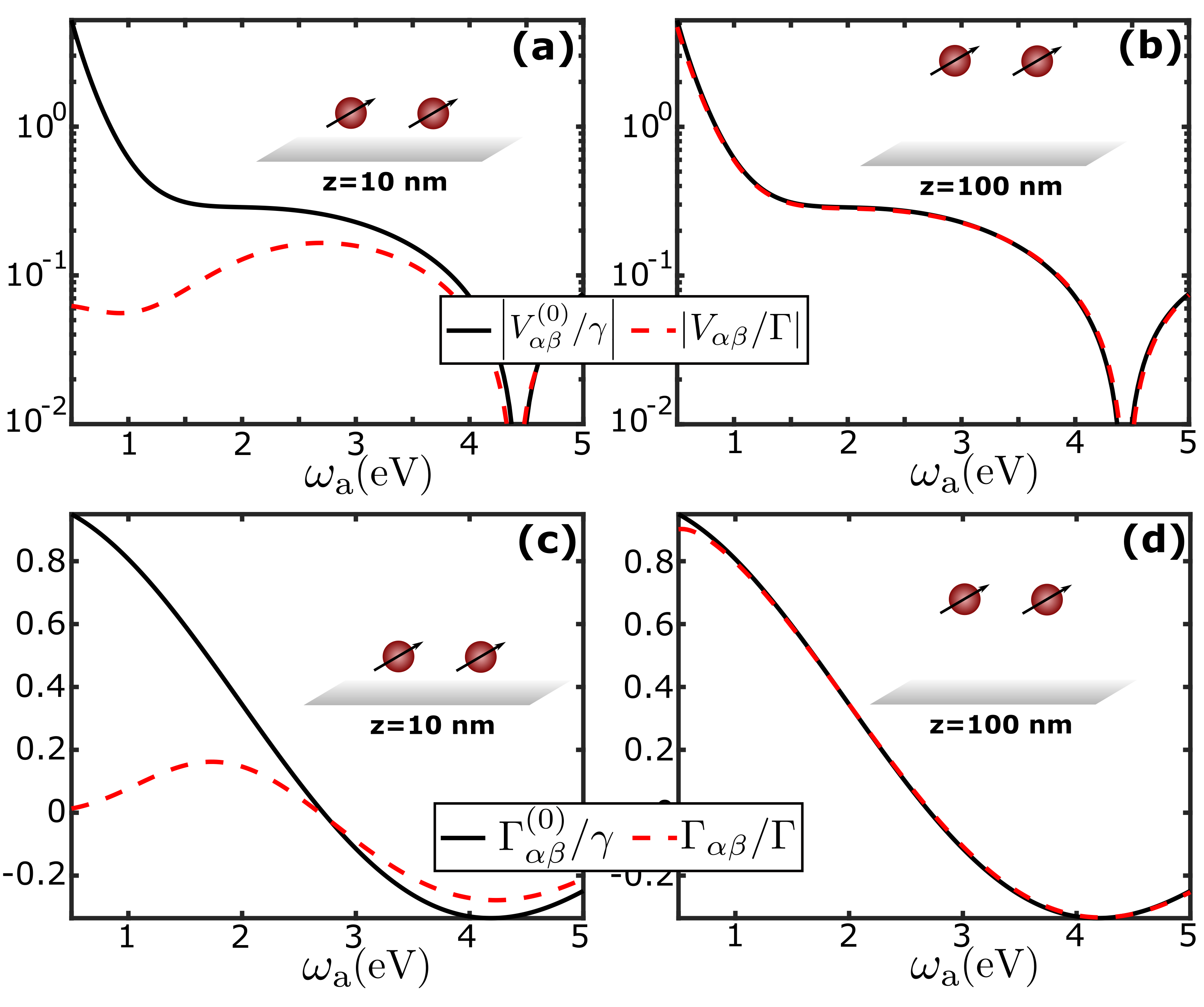}
\caption{Two dipoles at a fixed distance $a=200$ nm oriented parallel to each other and to the surface. All the quantities are plotted as a function of the transition frequency $\omega_{a}$. \textbf{(a)} and \textbf{(b)}: Coherent dipole-dipole interaction between two atoms in free-space, $V_{\alpha\beta}^{(0)}/\gamma$ (black solid line), and near a glass surface, $V_{\alpha\beta}/\Gamma$ (red dashed line). \textbf{(c)} and \textbf{(d)}: Off-diagonal element of the incoherent scattering matrix in free-space, $\Gamma_{\alpha\beta}^{(0)}/\gamma$ (black solid line), and in the vicinity of a surface, $\Gamma_{\alpha\beta}/\Gamma$ (red dashed line). The dipoles are at a distance $z=10$ nm for \textbf{(a)} and \textbf{(c)}, and $z=100$ nm for \textbf{(b)} and \textbf{(d)}.}\label{fig:Glass1D}
\end{figure}

The effect of the surface on the coherent and incoherent interactions in this case is fairly independent of the dipole orientation with respect to the surface (see Fig. \ref{fig:Glass1D} for the dipoles oriented parallel to the interface and each other). As in the case of silver, when the atoms are very close to the surface both coherent interactions $V_{\alpha\beta}/\Gamma$ and the off-diagonal elements of the dissipation coefficient matrix $\Gamma_{\alpha\beta}/\Gamma$ decrease dramatically. Hence, here the atoms decay independently at an enhanced rate $\Gamma$. However, while in the case of a metallic surface at larger distances the coherent interactions were enhanced, here the value of $V_{\alpha\beta}/\Gamma$ and $\Gamma_{\alpha\beta}/\Gamma$ simply approach the free-space ones as the distance from the surface is increased, so that for example at $z=100$ nm the effect on coherent and incoherent interactions is completely negligible.

\subsection{Atoms between two metallic surfaces}

\begin{figure}
\includegraphics[width=\columnwidth]{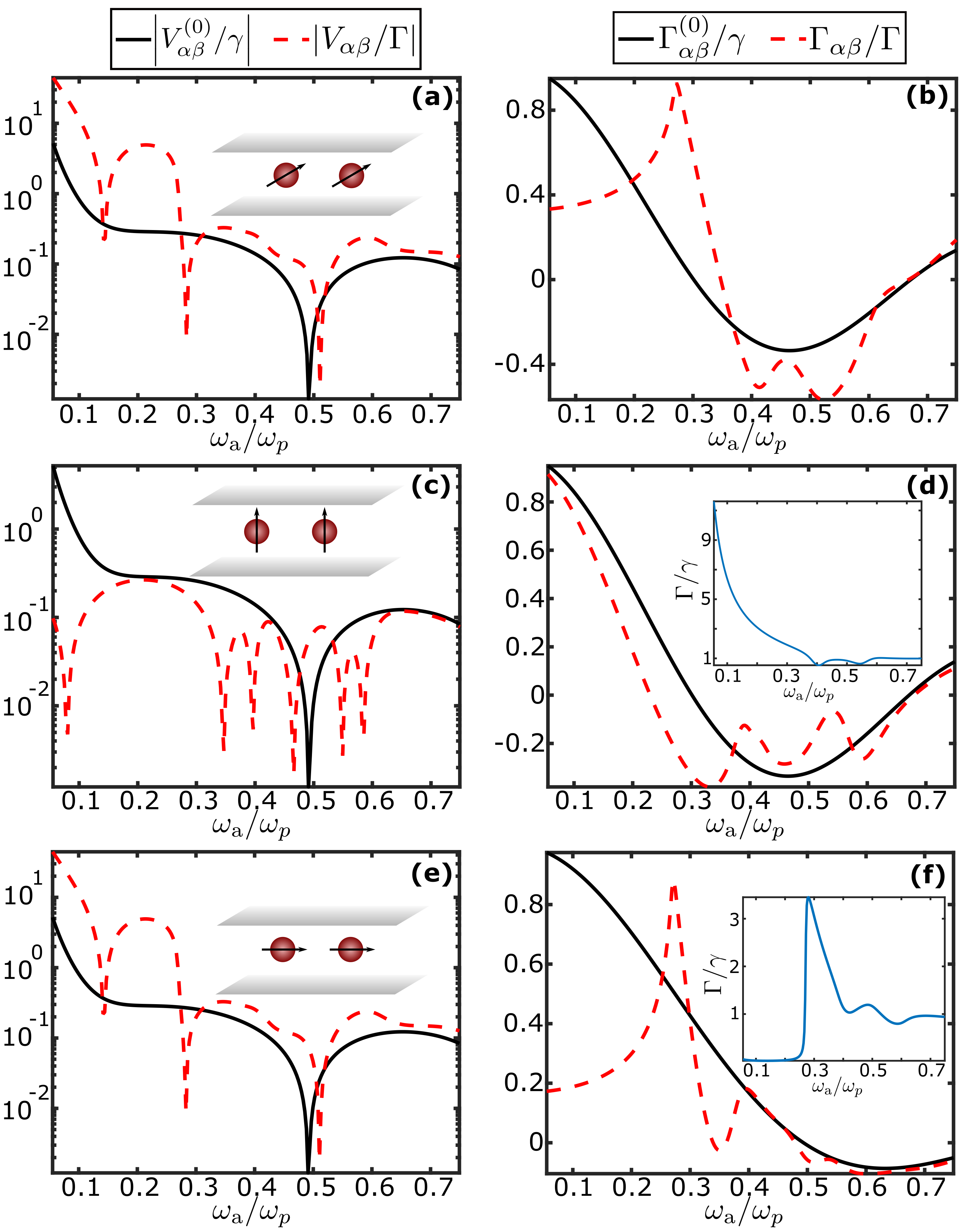}
\caption{Two dipoles at a fixed separation $a=200$ nm placed between two silver surfaces at a distance $z=100$ nm from each interface. All the quantities are plotted as a function of the transition frequency $\omega_{a}$ in units of the plasma frequency for Ag $\omega_p=9.01$ eV.  \textbf{(a)} and \textbf{(b)}: the dipoles are oriented parallel each other and to the surface. \textbf{(c)} and \textbf{(d)}: the dipoles are orthogonal to the surface. \textbf{(e)} and \textbf{(f)}: the dipole are aligned and parallel to the surface. The first column [\textbf{(a)}, \textbf{(c)} and \textbf{(e)}] shows the coherent dipole-dipole interaction in free-space, $V_{\alpha\beta}^{(0)}/\gamma$ (black solid line), and in the vicinity of a surface, $V_{\alpha\beta}/\Gamma$ (red dashed line). The second column [\textbf{(b)}, \textbf{(d)} and \textbf{(f)}] shows the off-diagonal element of the incoherent scattering matrix in free-space, $\Gamma_{\alpha\beta}^{(0)}/\gamma$ (black solid line), and in the vicinity of a surface, $\Gamma_{\alpha\beta}/\Gamma$ (red dashed line). The two insets show the modification of the single atom decay rate induced by the presence of the silver surface $\Gamma$, normalized by its free-space value $\gamma$ (dipole perpendicular \textbf{(d)} and parallel \textbf{(f)} to the surface).}\label{fig:VAg1D2S}
\end{figure}

The effects reported when the atoms are placed next to a surface can be enhanced by placing another surface forming the top layer [Fig. \ref{fig:system}(c) with $\varepsilon_{II}=\varepsilon_{I}$]. One can see an example of this enhancement in Fig. \ref{fig:VAg1D2S} for two atoms with $a=200$ nm between each other and placed at a distance $z=100$ nm both from a silver surface below and above them in the same three configurations of the dipoles studied before. We can see here [in (a), (c) and (e)] that interaction $V_{\alpha\beta}/\Gamma$ gets indeed enhanced with respect to the case of one surface [Fig. \ref{fig:VAg1D} (b), (d) and (f)], particularly for small values of the atomic frequency $\omega_\mathrm{a}/\omega_p$. The behavior of the coefficient $\Gamma_{\alpha\beta}/\Gamma$ changes much more dramatically, particularly for the cases where the dipoles are parallel to the two interfaces, with sudden changes at frequencies where the modified single atom decay rate $\Gamma$ also has sudden changes [see inset in Fig. \ref{fig:VAg1D2S}(f)]. Note that here, $\Gamma/\gamma$ tends to zero for $\omega_\mathrm{a}/\omega_p < 0.25$. The reason is that, at these frequencies, the size $h$ of the middle layer where the atoms sit is smaller than $\lambda/2$, and therefore the dipoles cannot couple to any radiative cavity modes \cite{dutra1996}. Hence, the decay rate is strongly suppressed as only non-radiative decay is possible.

\section{Application: excitation transport on a one-dimensional chain}

In this Section we will show in a specific example how the modification of the scattering properties due to the presence of a surface can be used for the improvement of the efficiency of the transport of an excitation in a one-dimensional lattice, increasing the lifetime of the excitation while preserving its transport due to the coherent dipole-dipole interactions.

\begin{figure}
\includegraphics[width=\columnwidth]{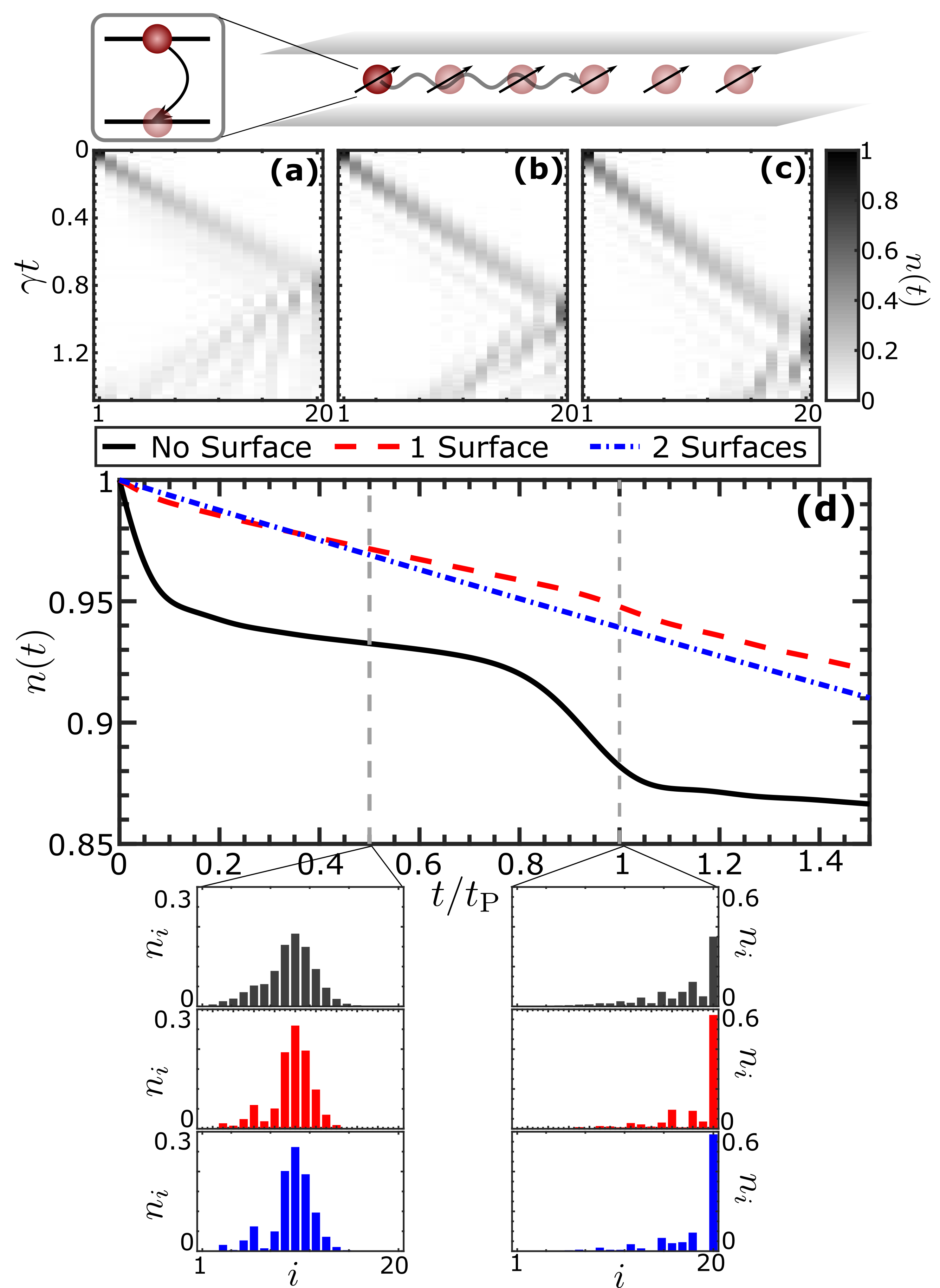}
\caption{A one-dimensional chain of 20 atoms where, initially, the leftmost atom is in the excited state while the rest are in the ground state. The population of the excited state in each site $i$, $n_i$ is plotted as a function of time for the following cases: \textbf{(a):} Atoms in free space; \textbf{(b):} At a distance $z=100$ nm from one silver surface; \textbf{(c):} The chain is at equidistance ($z=100$ nm) from two silver surfaces. Panel \textbf{(d):} The total excitation $n(t)$ decays as a function of time. The presence of a nearby silver surface notably suppresses the action of the dissipation, making the excitation longer lived. The insets at the bottom of the figure describe the excitation probability $n_i$ of each site $i$ on the lattice as the excitation has been transported to the other side of the lattice ($t=t_P$) and while still traveling ($t=t_P/2$).}\label{fig:Chain}
\end{figure}

The specific system we consider is a one-dimensional lattice with a lattice constant $a=206.4$ nm filled with identical strontium atoms, as it was proposed in \cite{olmos2013}. The two relevant energy levels in each atom are the triplet states ${}^{3}P_0$ and ${}^{3}D_1$, which represent the ground $\left|g\right>$ and excited $\left|e\right>$ states, respectively (note that we only consider here one of the three states that form the ${}^{3}D_1$ manifold, a situation that can be achieved, for example, by applying an external magnetic field to lift the degeneracy of the levels). The transition between them has an unusually long wavelength $\lambda=2.6$ $\mu$m, which corresponds to an atomic frequency $\omega_\mathrm{a}=0.48$ eV. We consider the many-body initial state to be the one where the leftmost atom is in the excited state and the rest in the ground state (in the case of Fig. \ref{fig:Chain} we consider 20 atoms). As the system evolves in time under the dynamics determined by the master equation in Eq. (\ref{MasterEq}), the coherent dipole-dipole interaction transports the excitation to the right while the action of the dissipation eventually will lead to the emission of a photon to the radiation field. For simplicity we consider that all dipole moments are parallel and aligned with the interface.

Let us consider first the free-space case [Fig. \ref{fig:Chain}(a) and (d)]. In \ref{fig:Chain}(a) we depict the value of $n_i\equiv\left<\hat{n}_i\right>$ with $\hat{n}_i=\left|e\right>_i\left<e\right|$ being the population of the excited state in the $i$-th atom, as a function of time. We observe that the excitation stays relatively localized and gets transported via the coherent interactions to the other side of the lattice. Since the dissipation has collective character, the presence of the rest of atoms makes the excitation much longer lived than if the excitation was placed in a single isolated atom (subradiant behavior), as it can be observed in Fig. \ref{fig:Chain} (d) (solid black line) where the total excitation $n(t)=\sum_i n_i$ is depicted. In particular, at very short times there is a drop of $n(t)$ followed by a plateau that lasts while the excitation wavepacket is in the bulk of the lattice, corresponding to the excitation occupying subradiant emission modes.

When the chain is near a silver surface ($z=100$ nm and $\omega_\mathrm{a}/\omega_p\approx0.05$), as we saw in Fig. \ref{fig:VAg1D}(b), the ratio between the coherent interaction and the decay, represented by $V_{\alpha\beta}/\Gamma$, is enhanced with respect to the value in free space. We also know from Fig. \ref{fig:GAg1D}(b) that the dissipation seems to keep at least partly its collective character, as $\Gamma_{\alpha\beta}\approx\Gamma_{\alpha\beta}^{(0)}$. This is indeed observed in Fig. \ref{fig:Chain}(b) and (d): the excitation is transported more efficiently to the other side of the lattice than in (a), where we consider the atoms in free space. In the case where the atoms are placed between two surfaces [Fig. \ref{fig:Chain}(c) and (d)] it is interesting to observe that the decay, while as slow as the one-surface case, does not have a collective character, as $n(t)$ follows an exponential decay and does not feature a plateau as in the surface-free case. Here, it is only the fact that $\Gamma\ll\gamma$ that makes the excitation longer lived.

Note that in \ref{fig:Chain}(d) we have introduced $t_P$ as the time when the amount of excitation on the rightmost site of the lattice reaches a maximum, as an approximate idea of how fast the excitation is transported across the lattice. Its value without a surface [Fig. \ref{fig:Chain}(a), $t_P=0.82\gamma$], is slightly shorter than in the presence of one surface [Fig. \ref{fig:Chain}(b), $t_P=0.97\gamma$] and with two surfaces [Fig. \ref{fig:Chain}(c), $t_P=1.15\gamma$].

\section{Conclusions}

In this paper, we have performed a detailed study on how the presence of a metallic or dielectric surface(s) in the vicinity of a dense atomic gas can modify some of its properties, in particular the coherent induced exchange interactions and the incoherent emission of photons to the radiation field. Specifically, we were interested to investigate how the interaction between the identical atoms and the surface excitations (e.g. surface plasmon polaritons) can modify possible collective atomic behaviors, which has already been investigated for atoms in free space. The effect of the surface(s) was included in the quantum master equation describing the behavior of the atomic system using the Green tensor formalism.

We have considered two different geometrical configurations, where the atoms are placed near a single surface or between two identical ones. We consider both metallic and dielectric materials to analyze the relevance of the surface plasmon polaritons occurring in metallic surfaces. All the expressions appearing in the master equation and characterizing the dynamics of the system were thoroughly investigated as a function of the system's parameters (atomic transition frequency, atom-surface separation, etc). Our analysis pointed out that, despite the fact that the surface can strongly modify the relevant parameters of the atoms' dynamics, e.g. enhancing the atoms' dipole-dipole exchange of excitations, if the interaction between the single atom and the surface is too strong with respect to the interatomic cooperativity, collective phenomena are suppressed rather than enhanced. In particular, for the metals considered in our investigation, for which surface plasmons can enhance the interatomic interaction, the dynamics of the total atomic system results from a balance among all the processes at work and strongly depend on the system's parameters. An enhancement of the system’s properties was particularly evident from the analysis of the energy transport along a one-dimensional atomic chain.

An exciting opportunity for further study is offered by different types of materials. Recent developments in the field of plasmonics have lead to the creation of materials which have significantly lower plasma frequencies than metals. For example, films of transparent conducting oxides such as tin oxide doped with indium or fluorine have been shown to have plasma frequencies in the near infrared \cite{lenski2009,Dominici2009}. Similarly doped semiconductors such as ZnO:Ga \cite{Sadofev13,Kalusniak14} have shown similar properties with very good figures of merit. These materials could enhance the coherent interatomic interaction with respect to the dissipation over a large range of parameters. This is particularly interesting for the strontium lattice system that we consider in Section IV, as the atomic transition is now much closer to the plasmon resonance frequency.

\begin{acknowledgements}
I.L. gratefully acknowledges funding through the Royal Society Wolfson Research Merit Award and funding  from the European Research Council under the European Union’s Seventh Framework Programme (FP/2007-2013)/ERC Grant Agreement No. 335266 (ESCQUMA). F.I. acknowledges support from the DFG through the DIP program (Grant No. SCHM 1049/7-1) and by the Deutsche Forschungsgemeinschaft (DFG) through project B10 within the Collaborative Research Center (CRC) 951 Hybrid Inorganic/Organic Systems for Opto-Electronics (HIOS). B.O. was supported by the Royal Society and EPSRC grant DH130145.
\end{acknowledgements}

\appendix

\section{Single atom frequency shifts}\label{app:shift}

In this Appendix, we discuss the presence of shifts in the transition frequencies of the atoms induced by the presence of a nearby surface. This shift depends on the distance from the surface and can be related to the van der Waals-Casimir-Polder interaction between an atom and a surface \cite{intravaia2011}. Here, we provide a calculation of the strength of these shifts in our system in order to justify the approximation used in the derivation of the master equation.

If we ignore the vacuum Lamb shift (as this is often absorbed into $\omega$) and consider only the effect of the surface(s), the shift is approximately given by
\begin{equation}
\delta = 3 \pi c \gamma \frac{\tilde{\omega}^2}{\omega^3} \hat{d}^* \mathrm{Re}\left[\underline{G}^R(\mathbf{r}, \mathbf{r}, \tilde{\omega})\right] \hat{d},
\end{equation}
with $\tilde{\omega} = \omega - \delta$ \cite{dung2-2002}. As $\delta$ is a function of the shifted frequency $\tilde{\omega}$, the relation between $\omega$ and $\tilde{\omega}$ is actually a self-consistent equation. We use an iterative method to determine $\tilde{\omega}$ for a given $\omega$, which then allows us to calculate $\delta$.

\begin{figure}[h]
\includegraphics[width=\columnwidth]{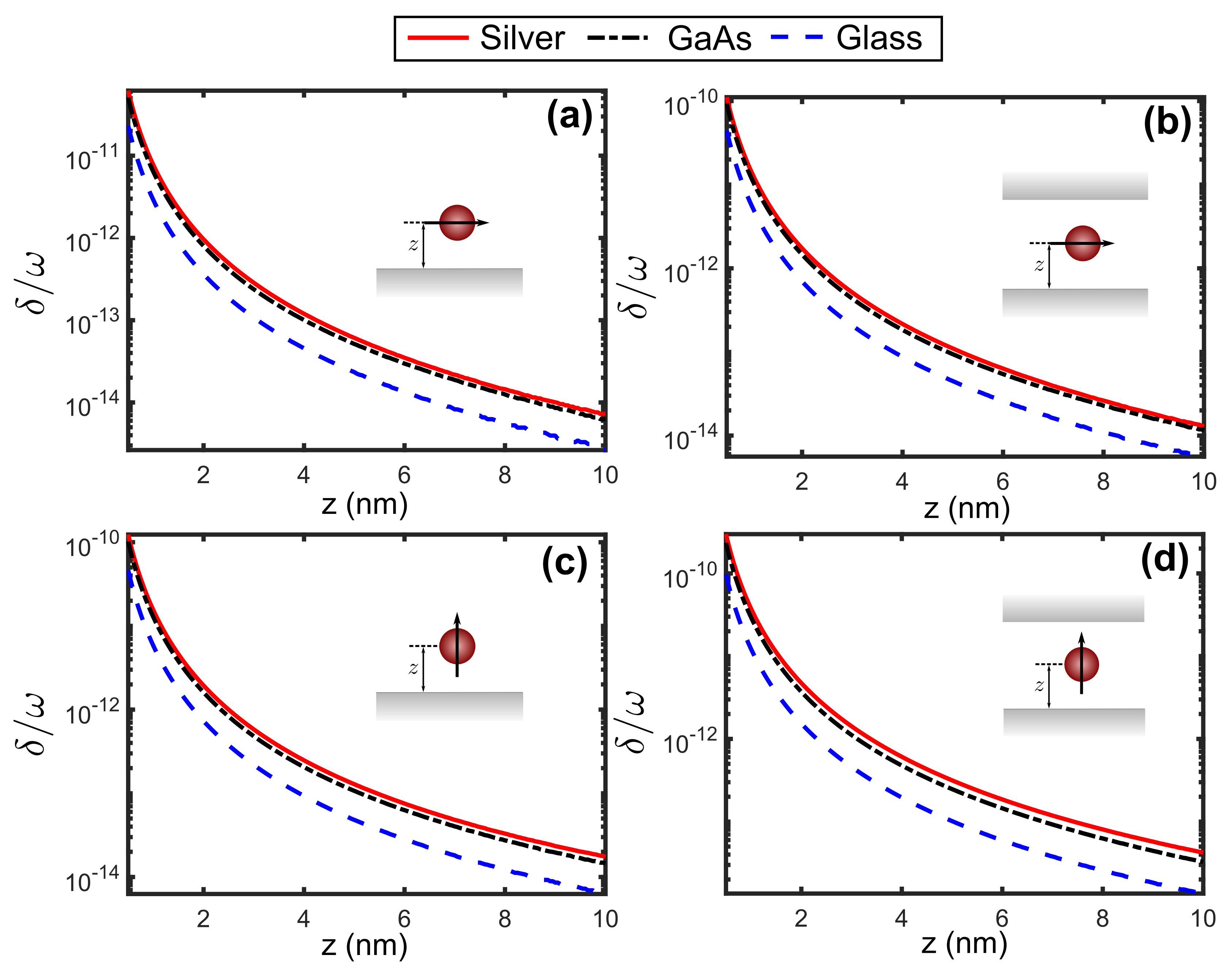}
\caption{Surface-induced frequency shift $\delta/\omega$ of a Sr atom ($\omega_{a}\approx$0.5 eV), plotted as a function of the atom-surface separation. The shift is calculated for metal (Ag shown) and dielectric (GaAs and glass) materials, in the case of both a single interface, and a three-layered system with $h = 2z$.} \label{fig:appendixfig1}
\end{figure}

Figure \ref{fig:appendixfig1} shows how $\delta$ varies with the atom-surface separation $z$ for a $\omega_\mathrm{a}\approx0.5$ eV transition (corresponding to the Sr case described in the main text) in the presence of one and two surfaces, for the materials considered in the main text (Ag and SiO${}_{2}$) and also GaAs for illustration purposes. We show this case as it is where we expect the shift to be largest, as it is when $k_\mathrm{a}z$ is smallest. The magnitude of the shift is many orders of magnitude smaller than $\omega$, even for the smallest atom-surface distances that we consider in this work ($z = 10$ nm). It is therefore safe for our purposes to consider the approximation $\tilde{\omega} = \omega$ in the master equation. Figures \ref{fig:appendixfig1}(a) and \ref{fig:appendixfig1}(b) depict the results for the atomic dipole parallel and \ref{fig:appendixfig1}(c) and \ref{fig:appendixfig1}(d) perpendicular to the plane of the surface(s), respectively. We also calculated the same values for gold and titanium, but these are omitted, as they are almost identical to the results obtained for silver.

\section{The expression for the scattering Green tensor}\label{scatteringtensor}

The scattering Green tensor can be written as
\begin{equation}
\underline{G}^R(\mathbf{r}_\alpha, \mathbf{r}_\beta, \omega) \!=\! \frac{i}{4 \pi}\!\! \int^\infty_0\!\!\!\! \mathrm{d}k_\rho \frac{k_\rho}{k_{z}} e^{i k_{z} h} \left(\underline{G}^s -  \frac{k^2_{z}}{k^2} \underline{G}^p\right).
\label{green tensor}
\end{equation}
Here, $k_{z} = \sqrt{\omega^2/c^{2} - k^2_\rho}$ and $k_\rho=\sqrt{k_{x}^{2}+k_{y}^{2}}$, i.e. the component of the wave-vector parallel to the surface.
In cartesian coordinates, the diagonal components of each of the two tensors $\underline{G}^s$ and $\underline{G}^p$ are given by
\begin{align}\label{eq:ReflGTcomp}
&G^s_{xx}=\frac{A_+^s}{2}\left[J_0(k_\rho \rho)+J_2(k_\rho \rho) \cos{2\phi}\right] \nonumber \\
&G^s_{yy}=\frac{A_+^s}{2}\left[J_0(k_\rho \rho)-J_2(k_\rho \rho) \cos{2\phi}\right] \nonumber \\
&G^s_{zz}=0 \\
&G^p_{xx}=\frac{A_-^p}{2}\left[J_0(k_\rho \rho)-J_2(k_\rho \rho) \cos{2\phi}\right]  \nonumber \\
&G^p_{yy}=\frac{A_-^p}{2}\left[J_0(k_\rho \rho)+J_2(k_\rho \rho) \cos{2\phi}\right] \nonumber \\
&G^p_{zz}=-\frac{k^2_\rho}{k^2_{z}}A_+^pJ_0(k_\rho \rho), \nonumber
\end{align}
while the off-diagonal ones read
\begin{align}\label{eq:ReflGTcomp}
&G^s_{xy}=-\frac{A_+^s}{2}J_2(k_\rho \rho) \sin{2\phi} \nonumber \\
&G^s_{yx}=G^s_{xy} \nonumber\\
&G^s_{xz}=G^s_{zx}=G^s_{yz}=G^s_{zy}=0 \nonumber \\
&G^p_{xy}=\frac{A_-^p}{2}J_2(k_\rho \rho) \sin{2\phi} \nonumber \\
&G^p_{xz}=\frac{ik_\rho}{k_{z}}B_+^pJ_1(k_\rho \rho) \cos{\phi} \\
&G^p_{yx}=G^p_{xy} \nonumber \\
&G^p_{yz}=\frac{ik_\rho}{k_{z}}B_+^pJ_1(k_\rho \rho) \sin{\phi} \nonumber \\
&G^p_{zx}=-\frac{ik_\rho}{k_{z}}B_-^pJ_1(k_\rho \rho) \cos{\phi} \nonumber \\
&G^p_{zy}=-\frac{ik_\rho}{k_{z}}B_+^pJ_1(k_\rho \rho) \sin{\phi}. \nonumber
\end{align}
In the previous expressions $\rho = \sqrt{x_{\alpha\beta}^2 + y_{\alpha\beta}^2}$, i.e. the distance between the atom in the plane parallel to the surface, $\phi = \cos^{-1}(x_{\alpha\beta} / \rho)$, and $J_n(x)$ is the Bessel function of order $n$. We have also defined the functions
\begin{eqnarray}\label{eq:CqDef}
A^q_{\pm}(k_\rho) &=& [r^q_- \mathrm{e}^{i k_{z} (z_\alpha + z_\beta - h)} + r^q_+ \mathrm{e}^{-i k_{z} (z_\alpha + z_\beta - h)} \nonumber \\
&&\pm 2 r^q_+ r^q_- \cos{(k_{z}z_{\alpha\beta})} \mathrm{e}^{i k_{z} h}]D^{-1}_q, \nonumber \\
B^q_{\pm}(k_\rho) &=& [r^q_- \mathrm{e}^{i k_{z} (z_\alpha + z_\beta - h)} + r^q_+ \mathrm{e}^{-i k_{z} (z_\alpha + z_\beta - h)}\nonumber  \\
&&\pm 2i r^q_+ r^q_- \sin{(k_{z}z_{\alpha\beta})} \mathrm{e}^{i k_{z} h}]D^{-1}_q,  \\
D_q(k_\rho) &=& 1 - r^q_+ r^q_- \mathrm{e}^{2i k_{z} h}. \nonumber
\end{eqnarray}
The functions $r^q_{\pm}$ are the Fresnel reflection coefficients defined in Eq.  \eqref{eq:FresnelC}.
The integral in Eq. \eqref{green tensor} must be evaluated numerically, and care must be taken in doing so, due to the potential presence of singularities in $k_{z}$ and the reflection coefficients. There are multiple approaches which can be used in order to avoid these singularities, e.g. by dividing the integral into the regions where $k_z$ is real or imaginary, and choosing a suitable change of integration variable in each region. In this work, we use the approach of Paulus et al. \cite{paulus2000} and deform the integration path in the complex plane of $k_\rho$, with an elliptical integration path used in the region of the singularities in order to avoid them.

\section{Other materials}\label{app:materials}

\begin{figure}[h]
\includegraphics[width=\columnwidth]{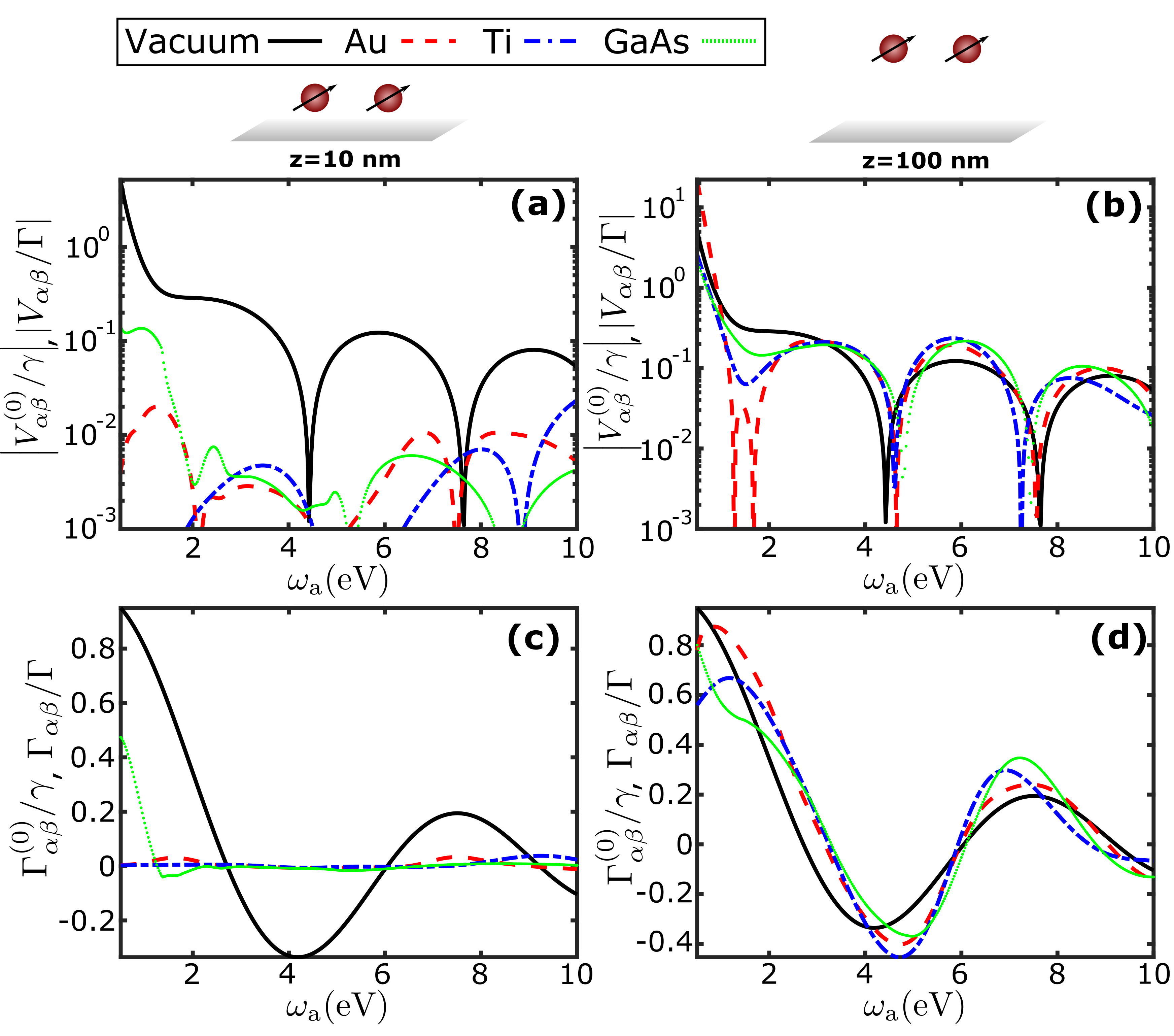}
\caption{Two dipoles at a fixed distance $a=200$ nm oriented parallel to each other and to the surface. All the quantities are plotted as a function of the transition frequency $\omega_{a}$. \textbf{(a)} and \textbf{(b)}: Coherent dipole-dipole interaction between two atoms in free-space, $V_{\alpha\beta}^{(0)}/\gamma$ (black solid line), and nearby surface of different materials, $V_{\alpha\beta}/\Gamma$. \textbf{(c)} and \textbf{(d)}: Off-diagonal element of the incoherent scattering matrix in free-space, $\Gamma_{\alpha\beta}^{(0)}/\gamma$ (black solid line), and in the vicinity of a surface, $\Gamma_{\alpha\beta}/\Gamma$. The dipoles are at a distance $z=10$ nm for \textbf{(a)} and \textbf{(c)}, and $z=100$ nm for \textbf{(b)} and \textbf{(d)}. \label{fig:appendixfig2}}
\end{figure}

For completeness, we show in this appendix some of the values for the modified coherent and incoherent interaction obtained when considering atoms near surfaces of different materials. In particular, we consider gold and titanium as examples of metals. We can see in Fig. \ref{fig:appendixfig2} (for atoms at interatomic distance $a=200$ nm) that indeed the results are very similar to the ones obtained in the main text for silver: for small distances to the surface, the interaction $V_{\alpha\beta}/\Gamma$ and the off-diagonal elements $\Gamma_{\alpha\beta}/\Gamma$ that represent the collective character of the dissipation are strongly suppressed. For larger distances, however, the value of the interactions for small values of the atomic frequency $\omega_\mathrm{a}$ can be increased with respect to the free-space ones, while $\Gamma_{\alpha\beta}/\Gamma$ remains very similar to $\Gamma^{(0)}_{\alpha\beta}/\gamma$.

We use the results described in \cite{rakic1996} for the modelling of the dielectric function of GaAs. The results for this surface are shown to be in some way in between the ones for metals and the ones for a glass surface shown in the main text.


%

\end{document}